\documentclass[paper]{JHEP3}
\pdfoutput=1
\usepackage{amsmath,amssymb,amsthm,amscd,graphicx}
\input epsf.sty

\theoremstyle{definition}

\newcommand{\CC}{{\cal C}}

\newcommand{\CN}{{\cal N}}
\newcommand{\CO}{{\cal O}}

\def\IZ{{\mathbb Z}}

\def\IC{{\mathbb C}}
\def\IP{{\mathbb P}}

\def\IS{{\mathbb S}}

\newcommand{\tr}{{\rm Tr}}
\newcommand{\re}{{\rm e}}
\newcommand{\ri}{{\rm i}}
\newcommand{\rd}{{\rm d}}

\newcommand{\sn}{{\rm {sn}}}
\newcommand{\cn}{{\rm {cn}}}
\newcommand{\dn}{{\rm {dn}}}
\newcommand{\cs}{{\rm {cs}}}

\newcommand{\kk}{K(k)}
\newcommand{\kkp}{K'(k)}
\newcommand{\cd}{{\rm cd }}
\newcommand{\nd}{{\rm nd }}
\newcommand{\dd}{{\rm d}}
\newcommand{\sd}{{\rm sd }}
\newcommand{\ssc}{{\rm sc }}

\newcommand{\be}{\begin{equation}}
\newcommand{\ee}{\end{equation}}
\newcommand{\ba}{\begin{aligned}}
\newcommand{\ea}{\end{aligned}}
\newcommand{\ben}{\begin{eqnarray}\displaystyle}
\newcommand{\een}{\end{eqnarray}}

\newcommand{\sectiono}[1]{\section{#1}\setcounter{equation}{0}}

\newdimen\tableauside\tableauside=1.0ex
\newdimen\tableaurule\tableaurule=0.4pt
\newdimen\tableaustep
\def\phantomhrule#1{\hbox{\vbox to0pt{\hrule height\tableaurule width#1\vss}}}
\def\phantomvrule#1{\vbox{\hbox to0pt{\vrule width\tableaurule height#1\hss}}}
\def\sqr{\vbox{%
  \phantomhrule\tableaustep
  \hbox{\phantomvrule\tableaustep\kern\tableaustep\phantomvrule\tableaustep}%
  \hbox{\vbox{\phantomhrule\tableauside}\kern-\tableaurule}}}
\def\squares#1{\hbox{\count0=#1\noindent\loop\sqr
  \advance\count0 by-1 \ifnum\count0>0\repeat}}
\def\tableau#1{\vcenter{\offinterlineskip
  \tableaustep=\tableauside\advance\tableaustep by-\tableaurule
  \kern\normallineskip\hbox
    {\kern\normallineskip\vbox
      {\gettableau#1 0 }%
     \kern\normallineskip\kern\tableaurule}%
  \kern\normallineskip\kern\tableaurule}}
\def\gettableau#1{\ifnum#1=0\let\next=\null\else
\squares{#1}\let\next=\gettableau\fi\next}

\newcommand{\figref}[1]{Fig.~\protect\ref{#1}}

\title{M-theoretic matrix models}

\author{
Alba Grassi and Marcos Mari\~no
\\
D\'epartement de Physique Th\'eorique et Section de Math\'ematiques,\\
Universit\'e de Gen\`eve, Gen\`eve, CH-1211 Switzerland\\
\\
\email{alba.grassi@unige.ch, marcos.marino@unige.ch}
}

\abstract{Some matrix models admit, on top of the usual 't Hooft expansion, an M-theory-like expansion, i.e. an 
expansion at large $N$ but where the rest of the parameters are fixed, instead of scaling with $N$. 
These models, which we call M-theoretic matrix models, appear in the localization of Chern--Simons--matter theories, and also in two-dimensional 
statistical physics. Generically, their partition function receives non-perturbative corrections which are not captured by the 't Hooft expansion. In this paper, we discuss 
general aspects of these type of matrix integrals and we analyze in detail two different examples. The first one is the matrix model 
computing the partition function of $\CN=4$ supersymmetric Yang--Mills theory in three dimensions with 
one adjoint hypermultiplet and $N_f$ fundamentals, which has a conjectured M-theory dual, and which we call the $N_f$ matrix model. 
The second one, which we call the polymer matrix model, 
computes form factors of the 2d Ising model and is related to the physics of 2d polymers. In both cases we determine their 
exact planar limit. In the $N_f$ matrix model, the planar free energy reproduces the expected behavior of the M-theory dual. 
We also study their M-theory expansion by using Fermi gas techniques, and we find non-perturbative corrections to the 't Hooft expansion.}

\begin{document}

\sectiono{Introduction}

In the last years, a new window has opened to understand the properties of M-theory and string theory on 
certain backgrounds: the combination of the AdS/CFT correspondence with supersymmetric localization in gauge theories. This combination has provided 
conjectural, exact results for quantities in M-theory, like for example Euclidean partition functions on certain AdS backgrounds, and 
it has led to new checks of the AdS/CFT correspondence. In the 
case of AdS$_4$, for example, the gauge theory computation of the Euclidean partition function 
on the three-sphere of Chern--Simons--matter theories reproduces at large $N$ the gravity calculation, including the $N^{3/2}$ behavior of membrane theories predicted in 
\cite{kt}. This was first done in \cite{dmp} in the case of ABJM theory \cite{abjm}, and it was extended to other models in many subsequent papers. 
One can even test AdS/CFT beyond leading order and match logarithmic corrections to the partition function \cite{bgms}. 

Perhaps the most interesting lesson for M-theory has been obtained in the study of non-perturbative corrections in ABJM theory. In this model, 
both worldsheet instanton corrections and membrane instanton corrections to the partition function can be computed in detail, by using the 
standard 't Hooft expansion \cite{dmp} and the 
so-called Fermi gas approach \cite{mp}, respectively. Due to a hidden (and perhaps accidental) connection 
to topological string theory \cite{mp1,hmo,hmmo, km}, 
one can obtain exact results for the {\it full} series  
of instanton corrections. A surprising aspect of this exact result is that the total contribution of 
worldsheet instantons (i.e. of fundamental strings) has infinitely many poles at physical values of 
the string coupling constant. These divergences are only cured if one adds membrane instantons and bound states of fundamental strings 
and membranes \cite{hmo,hmo2}. The cancellation 
of divergences in the partition function is known as the HMO mechanism. This mechanism 
shows, in a precise quantitative way, that a theory based solely on fundamental strings is radically incomplete, and a consistent theory is only obtained 
when one considers M-theory together with its solitonic objects, like membranes. 

From a more formal point of view, the results obtained for ABJM theory show that the 
't Hooft expansion does not capture the full physics of the model, since 
it only contains the contribution of fundamental strings. 
On general grounds, it has been known for a long time that 
the 't Hooft expansion is an asymptotic expansion, and in principle it has to be supplemented by non-perturbative contributions like large $N$ instantons 
(see \cite{mmreview} for a review of these issues). The study of the ABJM matrix model has 
shown that these corrections are not just a luxurious commodity: they are needed for 
consistency. 

These conclusions are probably generic for a wide class of AdS$_4$/CFT$_3$ duals, i.e. we expect that the 't Hooft expansion of the partition function of these models 
will miss an important part of the physics. Since the phenomena discovered in the study of the ABJM matrix model probe 
fundamental aspects of string theory (and of the large $N$ expansion), it is clearly important to study other examples where these aspects can be 
understood in detail, and where one can find exact results for the non-perturbative corrections.

In this paper we take some steps towards an understanding of what we call ``M-theoretic matrix models," 
i.e. matrix models which can be studied in both the 't Hooft expansion and in an M-theory expansion in which $N$ is large, 
but the coupling constants are kept fixed. 
The matrix models appearing in the localization of Chern--Simons--matter theories 
are of this type, as required by their duality with M-theory backgrounds, and their M-theory expansion was first considered in \cite{hkps}. 
There are other contexts in which similar models appear. For example, many matrix models considered in \cite{itep,kostov-rl}, 
which describe ADE models and their affine extensions on a random lattice, are also M-theoretic matrix models. In all these models, 
the 't Hooft expansion is likely to miss important ingredients, 
and it receives non-perturbative corrections which appear naturally in the M-theory expansion.

Unfortunately, the study of M-theoretic matrix models is difficult to pursue beyond ABJM theory, 
since in many cases we don't have a good control of their 't Hooft expansion, let alone 
of the non-perturbative corrections to it\footnote{ABJ theory \cite{abj} has been also extensively studied with similar 
techniques, see \cite{dmp,ahs,honda,mm}.}. As a matter of fact, even the planar limit of generic Chern--Simons--matter 
matrix models is difficult to obtain explicitly. Examples exist where 
this limit is more or less under control \cite{suyamaone,cmp,suyama, suyamatwo}, but the resulting 
expressions are often complicated and unilluminating. 

In this paper we analyze two matrix models whose planar limit can be determined exactly and is relatively simple. They 
might be exactly solvable in both, the 't Hooft expansion and the M-theory expansion, and they represent interesting laboratories 
to start the exploration of M-theoretic matrix models beyond ABJM theory. 
The first model, which we call the {\it $N_f$ matrix model}, 
calculates the partition function of a three-dimensional $\CN=4$ gauge theory 
which consists of an $U(N)$ vector multiplet coupled to one hypermultiplet in the adjoint representation and 
$N_f$ hypermultiplets in the fundamental. This theory is dual to M-theory on AdS$_4 \times \IS^7/N_f$, 
where the quotient by $N_f$ leads to 
an $A_{N_f-1}$ singularity \cite{bk,bcc, mp}. 
When $N_f=1$, this theory is equivalent to the ABJM theory with $k=1$, but for $N_f>1$ it describes a different M-theory 
background. In this paper we will solve the exact planar and genus one limit of this matrix model, which turns out to be 
relatively simple. This model can be also analyzed in the M-theory regime by using the Fermi gas approach. 
The perturbative grand potential was determined recently in \cite{mpufu}, and in this paper we give some results on its non-perturbative corrections. 
In particular, we find non-perturbative effects beyond the 't Hooft expansion, which are conjecturally due to membrane instantons, 
as in ABJM theory. 

The second matrix model that we study, which we call the {\it polymer matrix model}, is a particular case of the models considered in \cite{kostov-rl}, and it 
appears in the study of 2d polymers and in the calculation of correlation functions in the 2d Ising model. 
We solve exactly for its planar limit, and we also study it from the point of view of the 
Fermi gas, where it displays again non-perturbative effects which are not captured in the 't Hooft expansion. As an interesting bonus, we give a 
Fermi gas derivation of the function determining the short-distance behavior of the spin-spin correlation functions in the 2d Ising model. 
Both models, the $N_f$ matrix model and the polymer matrix model, can be regarded as particular cases of the $O(2)$ matrix model \cite{kostovon,ks}, 
and we use the technology developed in \cite{EK,EK2} to study their planar limit. They are also closely related to the models with adjoint multiplets 
studied in \cite{suyama}, but they turn out to be simpler. 

This paper is organized as follows: in section 2 we give a general overview of M-theoretic matrix models and their properties. In section 3 we study in detail the $N_f$ matrix model. We solve for 
its planar and genus one limit and we study it from the point of view of the Fermi gas. In section 4 we study the polymer matrix model using a similar approach. In section 5 we state some conclusions and 
prospects for future work. Appendix A contains some technical ingredients introduced in \cite{EK,EK2} to solve the $O(m)$ matrix model. Appendix B formulates the matrix models studied 
in this paper as Gaussian models perturbed by multi-trace potentials, and we explain a method to compute the relevant quantities at small 't Hooft coupling which can be used to check the exact 
solution. 

\sectiono{General aspects of M-theoretic matrix models}

In \cite{kwy, hama, jafferis}, explicit expressions in terms of matrix integrals were found for the partition functions on the three-sphere 
of various Chern--Simons--matter theories with $\CN\ge 2$ supersymmetry. 
The most studied example of this family of matrix models is ABJM theory \cite{abjm}. This is a quiver theory with two nodes, and 
each node is associated to a $U(N)$ Chern--Simons theory, with levels $k$ and $-k$, respectively. The partition function depends on $N$ and $k$ 
and it is given by the matrix integral, 
\be
\label{abjmmatrix}
\ba
&Z_{\rm ABJM}(N,k)\\
&={1\over N!^2} \int {\rd ^N \mu \over (2\pi)^N} {\rd ^N \nu \over (2\pi)^N} {\prod_{i<j} \left[ 2 \sinh \left( {\mu_i -\mu_j \over 2} \right)\right]^2
  \left[ 2 \sinh \left( {\nu_i -\nu_j \over 2} \right)\right]^2 \over \prod_{i,j} \left[ 2 \cosh \left( {\mu_i -\nu_j \over 2} \right)\right]^2 } 
  \exp \left[ {\ri k \over 4 \pi} \sum_{i=1}^N (\mu_i^2 -\nu_i^2) \right].
  \ea
  \ee
A natural generalization of this model is the family of necklace quivers constructed in \cite{jt,ik}. These theories are given by a
\be
U(N)_{k_1} \times U(N)_{k_2} \times \cdots U(N)_{k_r}
\ee
Chern--Simons quiver. Each node will be labelled with the letter $a=1, \cdots, r$. There are bifundamental chiral superfields $A_{a a+1}$, $B_{a a-1}$ connecting 
adjacent nodes, and in addition there can be $N_{f_a}$ matter superfields $(Q_a, \tilde Q_a)$ in each node, in the fundamental representation. We will write
\be
k_a=n_a k, 
\ee
and we will assume that 
\be
\label{add0}
\sum_{a=1}^r n_a=0. 
\ee
The matrix model computing the $\IS^3$ partition function of such a necklace quiver is given by 
\be
\label{necklace}
Z\left(N, n_a, N_{f_a}, k \right)={1\over (N!)^r} \int  \prod_{a,i} {\rd \lambda_{a,i} \over 2 \pi}  {\exp \left[ {\ri n_a k\over 4 \pi}\lambda_{a,i}^2 \right] \over \left( 2 \cosh{\lambda_{a,i} \over 2}\right)^{N_{f_a}} } \prod_{a=1}^r  {\prod_{i<j} \left[ 2 \sinh \left( {\lambda_{a,i} -\lambda_{a,j} \over 2} \right)\right]^2 \over \prod_{i,j} 2 \cosh \left( {\lambda_{a,i} -\lambda_{a+1,j} \over 2} \right)}.
\ee

These matrix integrals can be studied in two different regimes: in the 't Hooft expansion, one considers the limit 
\be
N, \,\, k, \, \, N_{f_a} \rightarrow \infty, 
\ee
but the 't Hooft and Veneziano parameters
\be
\lambda= {N \over k}, \qquad t_a={N_{f_a} \over k}
\ee
are fixed. In this regime, the free energy $F=\log Z$ has a $1/N$ expansion of the form 
\be
F\left(N, n_a, N_{f_a}, k\right)= \sum_{g\ge 0} k^{2-2g} F_g \left( \lambda, t_a, n_a\right).
\ee
This regime corresponds of course to the 't Hooft expansion of the original gauge theories, and to the genus expansion of the type IIA superstring duals. The 't Hooft expansion 
of these matrix models can be studied in principle by using standard large $N$ expansion techniques for matrix models. 

On the other hand, there is a M-theory expansion in which $N \rightarrow \infty$ but $k$ and $N_{f_a}$ are kept fixed. 
This makes contact with the M-theory dual and captures the strong coupling regime of the type IIA superstring. 
A study of the strict large $N$ limit of these models was first presented in \cite{hkps}. In \cite{mp} a different framework was proposed to study these models, 
based on the analogy between the matrix integrals (\ref{abjmmatrix}), (\ref{necklace}), and the canonical partition function $Z(N)$ of a one-dimensional Fermi gas. In this Fermi gas approach, 
all the information about the model is encoded in the spectrum of the one-particle Hamiltonian $\widehat H$ of the gas, or equivalently in the spectrum of 
an integral operator whose kernel is the density matrix: 
\be
\rho(x, x')= \langle x | \re^{- \widehat H} |x'\rangle.
\ee
 In this approach, a crucial r\^ole is 
played by the grand canonical partition function and the grand potential of the gas, which are defined as
\be
\Xi(z)= 1+ \sum_{N=1}^\infty Z\left(N \right) z^N, \qquad J(z)=\log\, \Xi(z). 
\ee
If one defines 
\be
\label{zell}
Z_\ell = \int \rd x_1 \cdots \rd x_\ell   \, \rho(x_1, x_2)\rho(x_2, x_3)\cdots \rho(x_{\ell-1}, x_\ell) \rho(x_{\ell}, x_1), 
\ee
then the grand potential can be computed as
\be
\label{jsum}
J(z)= -\sum_{\ell=1}^\infty {(-z)^\ell \over \ell} Z_\ell.
\ee

As an example of this formulation, let us quickly review the Fermi gas formulation of ABJM theory. 
The interaction term in the matrix integral (\ref{abjmmatrix}) can be rewritten by using the Cauchy identity, 
 \be
 \label{cauchy}
 \ba
  {\prod_{i<j}  \left[ 2 \sinh \left( {\mu_i -\mu_j \over 2}  \right)\right]
\left[ 2 \sinh \left( {\nu_i -\nu_j   \over 2} \right) \right] \over \prod_{i,j} 2 \cosh \left( {\mu_i -\nu_j \over 2} \right)}  
 & ={\rm det}_{ij} \, {1\over 2 \cosh\left( {\mu_i - \nu_j \over 2} \right)}\\
 &=\sum_{\sigma \in S_N} (-1)^{\epsilon(\sigma)} \prod_i {1\over 2 \cosh\left( {\mu_i - \nu_{\sigma(i)} \over 2} \right)}.
 \ea
  \ee
  In this equation, $S_N$ is the permutation group of $N$ elements, and $\epsilon(\sigma)$ is the signature of the permutation $\sigma$. After some manipulations, 
  one obtains \cite{kwytwo,mp}
\be
\label{fgasform}
Z_{\rm ABJM} (N,k)={1 \over N!} \sum_{\sigma  \in S_N} (-1)^{\epsilon(\sigma)}  \int  {\rd ^N x \over (2 \pi k)^N} {1\over  \prod_{i} 2 \cosh\left(  {x_i  \over 2}  \right)
2 \cosh\left( {x_i - x_{\sigma(i)} \over 2 k} \right)},
\ee
which can be immediately identified as the partition function of a one-dimensional ideal Fermi gas with density matrix
\be
\label{densitymat}
\rho_{\rm ABJM}(x_1, x_2)={1\over 2 \pi k} {1\over \left( 2 \cosh  {x_1 \over 2}  \right)^{1/2} }  {1\over \left( 2 \cosh {x_2  \over 2} \right)^{1/2} } {1\over 
2 \cosh\left( {x_1 - x_2\over 2 k} \right)}. 
\ee
Notice that, by using the Cauchy identity with $\mu_i=\nu_i$, we can rewrite (\ref{fgasform}) as
\be
\label{tanh-form}
Z_{\rm ABJM} (N,k)={1\over N!}  \int \prod_{i=1}^N {\rd x_i \over 4 \pi k}  {1\over 2 \cosh {x_i \over 2} } \prod_{i<j} \left( \tanh \left( {x_i - x_j \over 2 k } \right) \right)^2. 
\ee
The spectrum of the one-particle Hamiltonian is defined by the integral equation 
\be
\int \rho_{\rm ABJM}(x, x') \phi_n(x')\, \rd x = \re^{-E_n} \phi_n(x), 
\ee
where $\phi_n(x)$ are normalizable functions. The information about the large $N$ limit of the model can be recovered from the asymptotic behavior of the spectrum 
at large quantum numbers $n \gg 1$. It is easy to see from (\ref{densitymat}) that $k$ plays here the r\^ole of Planck's constant, therefore this asymptotic behavior 
can be obtained by adapting WKB techniques \cite{mp}, and one finds \cite{mp,hmo-exact,km}
\be
E_n^2 \approx { k \pi^2 \over 2} n, \qquad n \gg 1. 
\ee
It is easy to see that this behavior leads immediately to the $N^{3/2}$ behavior of the free energy predicted by classical supergravity \cite{kt}. 

A similar analysis can be 
made for the necklace quivers considered above. 
The asymptotic behavior of the energy levels is of the form $E_n^2 \approx n$ as in ABJM theory, but the precise coefficient 
(which was calculated in \cite{mp}) depends on the details of the quiver. The behavior of the energy levels at large $n$ leads to the following behavior at large chemical potential, 
\be
J(\mu) \approx {C \over 3} \mu^3, \qquad \mu \gg 1, 
\ee
where $C$ depends on the quiver. For example, for ABJM theory one finds 
\be
J(\mu)\approx {2 \mu^3 \over 3 \pi^2 k}, \qquad \mu \gg 1. 
\ee

The 't Hooft expansion and the M-theory expansion have been analyzed thoroughly only in the case of ABJM theory. 
The results of this analysis can be summarized as follows:

\begin{enumerate}

\item The free energy contains, at fixed $k$, a series of perturbative corrections in $1/N$, and on top of that a series of exponentially small corrections at large $N$, with an $N$-dependence of the 
form $\exp(-{\sqrt{N}})$. The perturbative corrections can be obtained either from the 't Hooft expansion \cite{dmp,dmp-np,fhm} or from the M-theory expansion \cite{mp}. 
They correspond conjecturally to perturbative quantum gravity corrections in M-theory \cite{dmp-np}, and the first, logarithmic correction, 
has been tested against a one-loop calculation in supergravity in \cite{bgms}. 

\item There are two types of non-perturbative corrections. Worldsheet instanton corrections are obtained naturally in the 't Hooft expansion, since they depend on $N$ through the 
't Hooft parameter. There are however exponentially small corrections at large $N$ which are non-perturbative in the string coupling constant, and are due to membrane instantons 
and bound states of membranes and fundamental strings. The pure membrane contribution can be in principle calculated in the Fermi gas approach \cite{mp,cm,km}, while bound 
states remain difficult to compute in both approaches \cite{hmo2,km}. The analytic calculations have been in addition tested against detailed numerical calculations \cite{py,hmo-exact,hmo}. 
The combination of all these approaches 
has led to a precise conjectural answer for the full series of non-perturbative corrections, which turn out to be determined by topological string theory and its refinement 
on a particular local Calabi--Yau manifold \cite{hmmo}. 

\end{enumerate}

The analysis of the ABJM matrix model shows that {\it the 't Hooft expansion is fundamentally incomplete}, 
since important non-perturbative effects can not be obtained in this framework. In fact, as noticed in \cite{hmo,hmo2}, the 't Hooft expansion leads to 
unphysical singularities in the free energy which need to be cured by the contribution of membranes and bound states (this is the HMO cancellation mechanism). 
From this point of view, the 't Hooft expansion is an inconsistent truncation of the theory. 
One of the advantages of the Fermi gas approach is that it gives an alternatively framework to analyze the large $N$ limit of these matrix models 
which captures some of these non-perturbative effects, and makes it possible to go beyond the 
't Hooft expansion. 

Although these results have been obtained for the ABJM model, we expect that similar features will appear in the 
the matrix models describing Chern--Simons--matter theories, i.e. we expect that 
the 't Hooft expansion of these models will miss important non-perturbative information. One reason to believe this is that these models admit an M-theory expansion 
where the large $N$ non-perturbative effects which are invisible in the 't Hooft expansion are no longer suppressed. 

In the case of Chern--Simons--matter models, the M-theory expansion is directly related to the existence of an M-theory dual. 
However, there are other matrix models which admit in a natural way an analogue of the M-theory expansion, in the sense that one can consider 
their behavior as $N$ becomes large but the rest of the parameters are fixed (instead of scaling with $N$, as in the 't Hooft limit). 
We will call these models ``M-theoretic matrix models." For example, the matrix models discussed in \cite{kostov-rl,itep}, as well as the matrix model of \cite{kkn}, 
are of this type. The examples considered in this paper are in fact particular 
examples of the $\widehat A_0$ matrix model of \cite{kostov-rl}: 
\be
\label{aohat}
Z_{\widehat A_0}(N,g_s)= {1\over N!} \int \prod_{i=1}^N {\rd z_i \over 2 \pi}   \, \re^{- {1\over g_s} V(z_i)} \prod_{i<j} { \left( z_i - z_j \right)^2 } \prod_{i,j} \left( z_i +z_j \right)^{-1}, 
 \ee
which is equivalent to the $O(2)$ matrix model \cite{kostovon,ks}. Another class of M-theoretic models are the matrix integrals computing Nekrasov's partition function. This partition function 
can be regarded as the grand canonical partition function of a classical gas \cite{ns}, and the number of particles is the instanton number of the underlying gauge theory.  In certain limits, 
this grand canonical partition function can be evaluated in closed form, as shown in \cite{ns,my,bourgine}. 

The M-theory expansion of a matrix model can be regarded as a direct thermodynamic limit, in which $N \rightarrow \infty$ 
but the other parameters are kept fixed. From the point of view of the 't Hooft expansion, this means that we consider a regime in which $N$ is large 
and the 't Hooft parameter scales with $N$. This regime has been also explored in \cite{afh}, where it is conjectured that, 
when both the 't Hooft limit and the M-theory limit 
exist, one can go from one to the other by an analytic continuation to strong coupling. In particular, \cite{afh} argue that planar dominance holds in the M-theory limit, 
i.e. that the M-theory limit of an amplitude is given by the continuation to strong 't Hooft coupling of its planar limit. We will see below that planar 
dominance holds in the examples we have studied. 

It is an interesting question to determine which matrix models admit a well-defined thermodynamic limit. 
We have not studied systematically this issue, but we can give some useful criteria: if a matrix model can be 
formulated as a quantum, one-dimensional, ideal Fermi gas (as proposed systematically in \cite{mp} but already pointed out in \cite{kostov-rl,kkn}), and if 
the resulting Hamiltonian has a discrete, 
infinite spectrum, then the thermodynamic limit exists and it is determined by conventional statistical-mechanical methods. Therefore, if a matrix model has a Fermi gas formulation, 
it definitely has an M-theory limit and it is an M-theoretic matrix model, as we have defined it. 

However, this criterion only applies to a restricted class of matrix models. What about a generic matrix model? Such a model, when written as an integral over 
the matrix eigenvalues, can be always regarded as a {\it classical} one-dimensional gas of $N$ particles. The existence of an M-theoretic thermodynamic limit 
will then depend crucially of the behavior of the potential and of the interaction terms. For example, if we regard the $\widehat A_0$ matrix model (\ref{aohat}) as the partition function of a classical 
gas, the interaction term decays at infinity, as expected from a conventional gas, and therefore we would expect it to have a good thermodynamic limit, as confirmed by the Fermi gas picture. 
In contrast, the standard Hermitian matrix model has 
an interaction term of the form $-\log|x| $. This does not decay at infinity and the existence of a good thermodynamic limit is not guaranteed. 
M-theoretic matrix models are excellent laboratories 
to understand the structure of non-perturbative effects in the large $N$ expansion, and also, via gauge/string dualities, non-perturbative effects in M-theory and string theory. We will 
now study in detail two examples which are relevant for AdS/CFT and for statistical physics in two dimensions, respectively.

\sectiono{The $N_f$ matrix model}

\subsection{Introducing the model} 

The theory we are going to consider is a supersymmetric $U(N)$, 
$\CN=4$ Yang--Mills theory in three dimensions, coupled to a single 
adjoint hypermultiplet and to $N_f$ fundamental hypermultiplets. When $N_f=1$, this theory is related by mirror symmetry to $\CN=8$ super Yang--Mills theory, therefore 
to ABJM theory with $k=1$ \cite{kwytwo}. From the point of view of M-theory, this gauge theory 
is supposed to describe $N$ M2 branes probing the space \cite{bk, bcc},
\be
\IC^2 \times \left( \IC^2/\IZ_{N_f} \right), 
\ee
where $\IZ_{N_f}$ acts on $\IC^2$ as
\be
\re^{2 \pi \ri/N_f}\cdot(a, b)= \left( \re^{2 \pi \ri /N_f} a, \re^{-2 \pi \ri /N_f} b\right). 
\ee
The corresponding quotient is an $A_{N_f-1}$ singularity, which can be resolved to give a multi-Taub-NUT space, as expected from the engineering of the theory in terms of D6 branes. 
The large $N$ dual description of this theory is in terms of M-theory on AdS$_4\times \IS^7/\IZ_{N_f}$, where the action of $\IZ_{N_f}$ is the one inherited by the action on $\IC^2 \times \IC^2$. 
  
The standard rules for localization of Chern--Simons--matter theories \cite{kwy,hama,jafferis} imply that the partition function on the three-sphere $\IS^3$ is given by the matrix integral  
\be
\label{pf}
Z(N, N_f)={1\over N!}  \int \prod_{i=1}^N {\rd x_i \over 4 \pi}  {1\over \left( 2 \cosh {x_i \over 2} \right)^{N_f}} \prod_{i<j} \left( \tanh \left( {x_i - x_j \over 2} \right) \right)^2. 
\ee
The $\tanh$ interaction between the eigenvalues includes both a $\sinh$ factor due to the Yang--Mills vector multiplet, and a $1/\cosh$ due to the hypermultiplet in the 
adjoint representation. Notice that, when $N_f=1$, this model leads to the same matrix integral than ABJM theory with $k=1$, in the representation (\ref{tanh-form}). We will solve this model in 
the planar and genus one limit as an exact function of the 't Hooft parameter
\be
\label{thooft}
\lambda={N\over N_f}
\ee
where $N_f$ is the number of flavours. The contribution of the hypermultiplets can be regarded as a one-body potential of the form $V(x)/g_s$, where
\be
V(x)= \log \left( 2 \cosh {x\over 2} \right), \qquad  g_s= {1\over N_f}. 
\ee
By comparing to the standard form of matrix models, we see that $1/N_f$ plays the r\^ole of string coupling constant. 
After making the change of 
variables $z=\re^x$, the matrix integral (\ref{pf}) takes the form (\ref{aohat}), with a potential  
 \be
 V(z)= \log \left(z^{1/2} + z^{-1/2}\right). 
 \ee
 The model (\ref{pf}) can be regarded as a particular example of the $O(2)$ model (\ref{aohat}), which is in turn an special case 
 of the $O(m)$ matrix model first introduced in \cite{kostovon} and further studied in for example 
 \cite{ks,EK, EK2,be}. The general form of the $O(m)$ matrix model is 
 \be
 \label{genom}
 Z(N, g_s, m)= \int \prod_{i=1}^N \rd z_i  \, \re^{- {1\over g_s} V(z_i)} \prod_{i<j} { \left( z_i - z_j \right)^2 }\prod_{i,j} \left( z_i +z_j \right)^{-{m/ 2}}.
 \ee
 The solution of the $O(2)$ model in the planar  limit was worked out in \cite{ks,EK}, but it turns out that it is more efficient to consider first 
the solution for the general $O(m)$ model, and then take the limit $m \rightarrow 2$. The reason is that there is a formalism 
to solve the $O(m)$ model for generic $m$ \cite{EK2} which incorporates in an efficient way the elliptic geometry of the planar solution. We will now solve the 
planar limit of the $N_f$ matrix model. 

\subsection{The planar solution}
In the approach of \cite{EK,EK2}, in order to solve the model (\ref{genom}), 
one introduces a planar resolvent in the standard way, 
\be
\omega_0(p)= \lim_{N \rightarrow \infty} {1\over N} \left\langle \tr {1\over p-M}\right\rangle,  
\ee
where $p$ is an exponentiated variable which lives in the $z$-plane. In terms of the density of eigenvalues $\rho(p)$, this reads
\be
\label{planar-res}
\omega_0(p)= \int \rd z {\rho(z) \over p-z}. 
\ee
 We will assume that our solution has one single cut in the $z$-plane, located at $[a,b]$. One important ingredient of the solution is of course 
 to find the relationship between the endpoints of the cut and the 't Hooft parameter. The solution of the planar limit of the model is encoded in 
 an auxiliary function $G(z)$, which was defined and used in \cite{EK} to solve the $O(n)$ model. It was determined explicitly in terms of theta functions in 
 \cite{EK2}. Since this function will play an important r\^ole in the solution of the model, we list its most important properties in Appendix \ref{gfunction}.
  It depends on a parameter $\nu$, which is in turn related to $m$ through the equation
 \be
 m =2 \cos(\pi \nu).
 \ee
Notice that the limit $m\rightarrow 2$ in which we are interested corresponds, in terms of this variable, to 
the limit $\nu \rightarrow 0$. According to the results of \cite{EK2}, the endpoints of the cut are determined by the two equations
\footnote{On the r.h.s of the second equation there is an overall factor of 1/2 w.r.t. the conventions in  \cite{EK2}.}
 \be
 \label{o2-end}
 \ba
 M_0 &={1\over 2 \cos {\pi (1-\nu) \over 2} }  \oint_{C} {\rd z \over 2 \pi \ri} V'(z)  G^{(1-\nu)} (z)=0,\\
 M_{-1}&={1\over 2 \cos {\pi \nu \over 2} } \oint_{C} {\rd z \over 2 \pi \ri} z V'(z)  G^{(\nu )}(z)={1\over 2}(2-m)\lambda. 
 \ea
 \ee
 where $\lambda=g_s N$ is the 't Hooft parameter of the model, and $\CC$ is a contour encircling the cut $[a,b]$. The 
 indices  $\nu$, $1-\nu$ indicate that the function $G$ should be evaluated for these values of the parameter. 
 These equations generalize the standard conditions determining the endpoints of the cut for the Hermitian one-matrix model.  
 Once the endpoints of the cut have been determined, one should calculate the planar free energy. Our convention for the genus 
 expansion of the free energy is 
\be
\label{ge-nf}
F(N, N_f)= \sum_{g\ge0} g_s^{2g-2} F_g(\lambda). 
\ee
A useful result in \cite{EK2} expresses the third derivative of the planar free energy w.r.t. the 't Hooft parameter, in terms of the endpoints of the cut $a,b$:
 \be
 \label{d3f}
 {\rd^3 F_0 \over \rd \lambda^3}= \left( 1-{m\over 2} \right) {1\over b^2-a^2} \left( {e^2 -a^2 \over a^2} {\rd a^2 \over \rd \lambda} -{e^2 - b^2 \over b^2}  {\rd b^2 \over \rd \lambda}  \right). 
 \ee
 In this equation, $e$ is a function of $a$, $b$ and $\nu$ defined in (\ref{ef}). This expression has a well-defined limit for $m\to 2$, which corresponds to taking 
 $\nu \rightarrow 0$. In this limit, the prefactor goes to zero, but $e^2$ diverges, as shown in (\ref{el0}). Therefore, only the terms proportional to $e^2$ (\ref{d3f}) survive, 
 and one obtains a finite result, 
 \be
 \label{free0}
 {\rd^3 F_0 \over \rd \lambda^3}=- {\pi^2 \over 2(b^2-a^2) k^2 \left(K' (k)\right)^2 } \left( {\rd a^2 \over \rd \lambda} -{a^2 \over b^2}  {\rd b^2 \over \rd \lambda} \right),
 \ee
where $K'(k)= K(k')$ is the elliptic integral of the first kind, with $k^2+ (k')^2=1$, and $k=a/b$, as in equation (\ref{kab}) of the Appendix \ref{gfunction}. It is possible to integrate this once w.r.t. $\lambda$ 
to obtain, 
\be
\label{freetau}
{\rd^2 F_0 \over \rd \lambda^2}=-2\pi {K(k) \over K'(k)}+ {\rm constant}= -{2 \pi \ri \over \tau} + {\rm constant}, 
\ee
where $\tau$ is given in (\ref{tauK}).
 
In the matrix model corresponding to the $N_f$ model, one has, 
 \be
 V'(z)= {1\over 2 z} {z-1 \over z+1},
 \ee
and we can calculate $M_0$ and $M_{-1}$ by residue calculus. One obtains,  
 \be
 \label{ab-first}
 \ba
{1\over 2 \cos {\pi (1-\nu) \over 2} } \left( 2 G^{(1-\nu)} (-1) -G^{(1-\nu)} (0) \right)&=0,  \\
{1\over 2 \cos {\pi \nu \over 2} } \left( -G^{(\nu)} (-1) +\cos\left( {\pi \nu \over 2}\right)\right) &= {1 \over 2} (2-m)\lambda.
\ea
\ee
These two equations have a non-trivial limit as $\nu \rightarrow 0$, which leads to the solution of the model.

We will however analyze a slightly different set of equations which were obtained by Suyama in a closely related context. In \cite{suyama}, the planar limit of 
supersymmetric Chern--Simons with $n$ adjoint multiplets was analyzed in detail, by using as well the correspondence with the $O(m)$ matrix model. However, the definitions of 
the resolvent and the map to the $O(m)$ model were slightly different from the ones explained above. To see how this goes, let us first extend our original matrix integral (\ref{pf}) to the 
case in which there are $n$ adjoint multiplets, 
\be
\label{pfn}
Z(N, N_f,n)={1\over N!}  \int \prod_{i=1}^N {\rd x_i \over 2 \pi}  {1\over \left( 2 \cosh {x_i \over 2} \right)^{N_f}} \prod_{i<j} { \left[2  \sinh \left( {x_i - x_j \over 2} \right) \right]^2 
\over  \left[ 2  \cosh \left( {x_i - x_j \over 2} \right) \right]^{2n} } . 
\ee
Let us then define the function 
\be
\label{v-fun}
v(z)= \lim_{N\to \infty} {\lambda \over N}\left\langle \sum_{i=1}^N  {z+ z_i \over z-z_i} \right \rangle. 
\ee
This function contains all the relevant information about the planar limit of the model. In fact, it is related to the standard resolvent, defined as in (\ref{planar-res}), 
by 
\be
v(z)= -\lambda + 2z \lambda \omega_0(z). 
\ee
In particular, it has the expansion at large $z$ given by 
\be
\label{largez-exp}
v(z) = \lambda + {2\lambda \langle W \rangle \over z} + \CO(z^{-2}), 
\ee
where
\be
\langle W \rangle=\lim_{N\to \infty} \left\langle \sum_{i=1}^N  z_i \right \rangle
\ee
is the VEV of a Wilson loop in the original gauge theory, in the fundamental representation. 

It is easy to derive the saddle-point equations for this matrix model and express them in terms of the function $v(z)$, 
\be
\label{sp-two}
{z-1 \over z+1} = v(z + \ri 0) + v(z-\ri 0) -2n v(-z), 
\ee
In order to use the formalism of $O(m)$ matrix models, one has to write
\be
n =-\cos(\pi \nu).
\ee
In this way the saddle-point equation (\ref{sp-two}) becomes identical to that of an $O(m)$ matrix model with $m=-2n$. In our original matrix model (\ref{pf}) we have $n=1$, 
so we have to consider now the limit $\nu \to 1$ of the results in \cite{suyama} (this is in contrast, and should not be confused, 
with the method explained above and derived directly from \cite{EK,EK2}, where one takes the limit $\nu \to 0$.) 

Let us then work out the planar solution of the model in detail. 
The first thing we should take into account is that in the original matrix integral, both the potential and the interaction are symmetric under $x\rightarrow -x$. 
Therefore, the density of eigenvalues is a symmetric function in the $x$ variable, and its support is of the form $[-A, A]$. In the $z$-plane the support is the interval $[a,b]$, 
and
\be
\label{Aa}
a=\re^{-A}, \quad b=\re^A
\ee
therefore
\be
b=1/a. 
\ee
The equation determining the 't Hooft parameter $\lambda$ as a function of $a$ can be deduced from the results in \cite{suyama}.  
In this paper, Suyama finds the solution of the saddle--point equation 
\be
\label{sp-gen}
V_\xi '(z)= v(z + \ri 0) + v(z-\ri 0) -2n v(-z), 
\ee
for a family of potentials of the form 
\be
V_\xi '(z)= -{2\over (\xi-1) z} {z-1 \over z- \xi}
\ee
and arbitrary $n$. In \cite{suyama}, these potentials were introduced as auxiliary objects which, after integrating w.r.t. $\xi$, 
lead to the logarithmic squared potentials typical of Chern--Simons matrix models \cite{mmcs,tierz}. Our case (\ref{sp-two}) is recovered by simply setting $\xi=-1$, and 
one finds from \cite{suyama} 
\be
\label{venp}
 \lambda= -{1\over n-1} \left[ {\ri \over 2 e \sin {\pi \nu \over 2}} G(-1) +{1\over 2} \right], 
 \ee
where $e$ is again given in (\ref{ef}) and $G(-1)$ is the function $G(z)$, evaluated at $z=-1$. 
It is not obvious that the above expression has a smooth limit when $n \rightarrow 1$, but this is the case, and in this limit (\ref{venp}) 
is equivalent to the limit $m\to 2$ of the second equation in (\ref{ab-first}). To determine analytically this limit, 
 we have to expand the function $G(z)$ around the point $\nu=1$.  After some calculations, which are sketched in the Appendix, 
 one finds the surprisingly simple equation 
 \be
 \label{tex}
 \lambda= -{1\over 8} + {(1+k)^2 \over 8 \pi^2} K'(k)^2, 
 \ee
 where the elliptic modulus is 
 \be
 k=a^2
 \ee
This of course is a particular case of (\ref{kab}) when $b=1/a$. 

The above equation determines the 't Hooft parameter as a function of the endpoint of the cut $a$. It is immediate to verify that, for $a=1$, $\lambda=0$, as it should. The free energy 
now follows from (\ref{free0}) and (\ref{tex}). One finds, 
 \be
 \label{three-F}
  {\rd^3 F_0 \over \rd \lambda^3}= {4 \pi^4 \over (1+a^2)^2\left(K'(k) \right)^3} {1\over E'(k)- a^2 K'(k)}.
  \ee

It is instructive to compare these results with a direct calculation of the endpoint of the cut and the free energy around $\lambda=0$, by treating (\ref{pf}) as a multi-trace matrix model. This perturbative method 
is explained in Appendix B, and leads to the following expansions:
\be
\ba
\label{pert-aF}
{A^2 \over 4} (\lambda)&=4 \lambda - {4 \lambda^2\over 3} + {392 \lambda^3 \over 45} - {4808 \lambda^4 \over 105} +\CO(\lambda^5), \\
F_0(\lambda)&= {\lambda^2 \over 2} \left( \log \lambda- {3\over 2} \right)-\log(2)\lambda-{\lambda^3\over 2} + {19 \lambda^4\over 24} - {9 \lambda^5\over 4}+ \CO( \lambda^6), 
\ea
\ee
where $A= -\log a$ is the endpoint of the cut in the $x$-plane, see (\ref{Aa}). 
The weak-coupling expansion, together with (\ref{three-F}), determines completely $F_0(\lambda)$. The integration constant in (\ref{freetau}) can be fixed by the second equation in (\ref{pert-aF}), and it 
turns out to be zero, therefore, 
  \be
  \label{free2}
  {\rd^2 F_0 \over \rd \lambda^2}= -{2\pi\ri \over \tau}=-2 \pi { K(a^2) \over K(\sqrt{1-a^4})}.
  \ee
  \subsection{The genus one free energy}
  The next-to-leading term in the 't Hooft expansion (\ref{ge-nf}) is the genus one free energy $F_1(\lambda)$. A general expression for this free energy in the $O(2)$ 
  matrix model has been found in 
  \cite{EK}\footnote{ The expression written down in \cite{EK} seems to have some misprints: the term $a^2/48$ should be $\log(a^2)/48$, and for general $a,b$, one should have $k_{a}=(1-a^2/b^2)^{1/2}$.}. In our case, this reads:
 \be\label{F1EK} F_1(\lambda)=-{1\over 24}\log(M_1 J_1)-{1\over 6} \log(1/a^2-a^2)-{1\over 4}\log(K(k_a)K(k_b)),
 \ee
 where 
 \be k_b^2=-1/a^4 + 1, \quad k_a^2=-a^4 + 1,
  \ee
   and $M_1$ and $J_1$ are moments given by contour integrals, which can be computed explicitly in terms of elliptic integrals of the first, second and third kinds, 
 \be
 \label{M1J1}
  \ba
   M_1=&\oint \limits_{[a,b]} {\dd z \over 2\pi \ri}{z-1\over 2(z+1)}{1\over (z^2-a^2)^{3/2}(z^2-a^{-2})^{1/2}}  \\
 =& \frac{1}{2a \left(-1+a^2\right)^2 \left(1+a^2\right) \pi } \Bigg\{-\left(1+a^2\right)^2 E\left[\frac{-1+a^2}{1+a^2}\right]+ \\
& 2a^2 \left(\left(a^2+2 a+3\right) K\left[\frac{-1+a^2}{1+a^2}\right]-4 a \Pi\left[\frac{(-1+a)^2}{1+a^2},\frac{-1+a^2}{1+a^2}\right]\right) \Bigg\},\\ 
 J_1=&\oint \limits_{[a,b]} {\dd z \over 2\pi \ri}{z-1\over 2(z+1)}{1\over (z^2-a^2)^{1/2}(z^2-a^{-2})^{3/2}} \\
 =&-\frac{1}{2 \left(-1+a^2\right)^2 \left(1+a^2\right) \pi }a^3 \Bigg\{\left(1+a^2\right)^2 E\left[\frac{-1+a^2}{1+a^2}\right]-
 \\
 &2 \left(\left(1+2 a+3 a^2\right) K\left[\frac{-1+a^2}{1+a^2}\right]-4 a \Pi\left[\frac{(-1+a)^2}{1+a^2},\frac{-1+a^2}{1+a^2}\right]\right)\Bigg\}.
 \ea\ee
The weak coupling expansion of (\ref{F1EK}) is given by
\be F_1(\lambda) 
=-\frac{\log (\lambda )}{12}-\frac{1}{6} \log \left(\frac{\pi ^3}{2}\right)+\frac{3 \lambda }{4}-\frac{19 \lambda ^2}{24}+\frac{25 \lambda ^3}{12}-\frac{271 \lambda ^4}{48}+\mathcal{O}(\lambda^5).\ee
This expansion matches with a direct perturbative computation in the matrix model.
  
\subsection{Resolvent, Wilson loop and density of eigenvalues}

The function $v(z)$, which contains all the information about planar correlators, can be also obtained from the results of \cite{suyama}. It has the form, 
\be
v(z)= {1 \over n^2-1} \left( f(z) +n f(-z) \right)+\omega(z) , 
\ee
where
\be
f(z)=-{1\over 2} {z-1\over z+1},
\ee
\be 
\omega(z)=- \ri \left(\re^{\ri \pi \nu /2} \omega_+(z) -\re^{-\ri \pi \nu /2}\omega_+(-z) \right),
\ee
and
\be
\omega_+(z)= {1\over 2(n^2-1) e} { \cn (u) \dn (u) \over z^2-1} G(-1) \left( z^{-1} G_+(z^{-1}) + z G_+(z) \right). 
\ee
The variable $u$ is related to $z$ through (\ref{u-def}).
The above expressions are obtained for generic $n$. We can now take the limit $n\rightarrow 1$. All the apparent divergences cancel, and we find the 
explicit expression
\be \ba
\label{vfinal} v(z)&={1\over 2 \pi ^2 z^2 \left(-1+z^2\right) }\left\{
\pi ^2 z^2 \left(-z+2 \lambda \left(1+z^2\right)\right)+a^2 \left(\kkp \right)^2 \right.\\
& \left.+z \left(\left(z+z^3\right) A(u) (\pi +A(u))-a \cn(u) \dn(u) (\pi +2 A(u)) \kkp \right. \right.\\
&\left.\left. -\left(1+a^2+a^4\right) z \left(\kkp \right)^2\right)\right\},
\ea
\ee
where 
\be
A(u)={\pi \over 2\kk}u +K'(k)  \left( {\vartheta'_1 \over \vartheta_1 } \right) \left({u\over 2\kk} \right). 
\ee
Notice that this function is not algebraic in $z$, in contrast to the resolvent of ABJM theory \cite{hy,dmp}. 
By using the expansion (\ref{largez-exp}), we can extract the exact value of the Wilson loop vev, 
\be 
\label{WL} 2\lambda \langle W\rangle =-\frac{1}{2}+\frac{1}{4 a \kk}+\frac{K'(k)}{2 a \pi }+\frac{aK'(k)}{2 \pi }-\frac{E(k) K'(k)}{2 a \kk \pi }.
\ee 
This can be expanded near $\lambda=0$, and one finds, 
\be 
\langle W\rangle=1+2\lambda +4 \lambda^3+\mathcal{O} (\lambda^{4}), 
\ee
which agrees with an explicit perturbative computation. 

\begin{figure} \begin{center}
 {\includegraphics[scale=0.5]{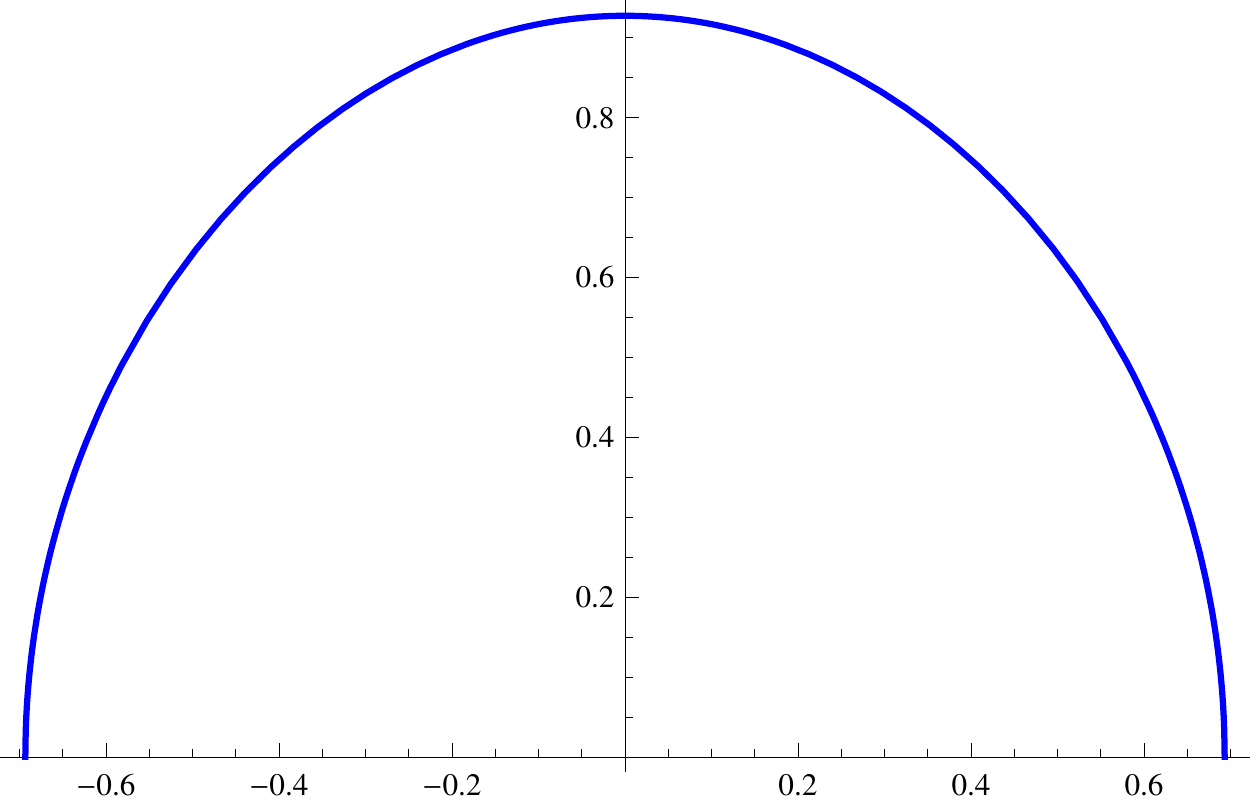} \qquad  \qquad \includegraphics[scale=0.5]{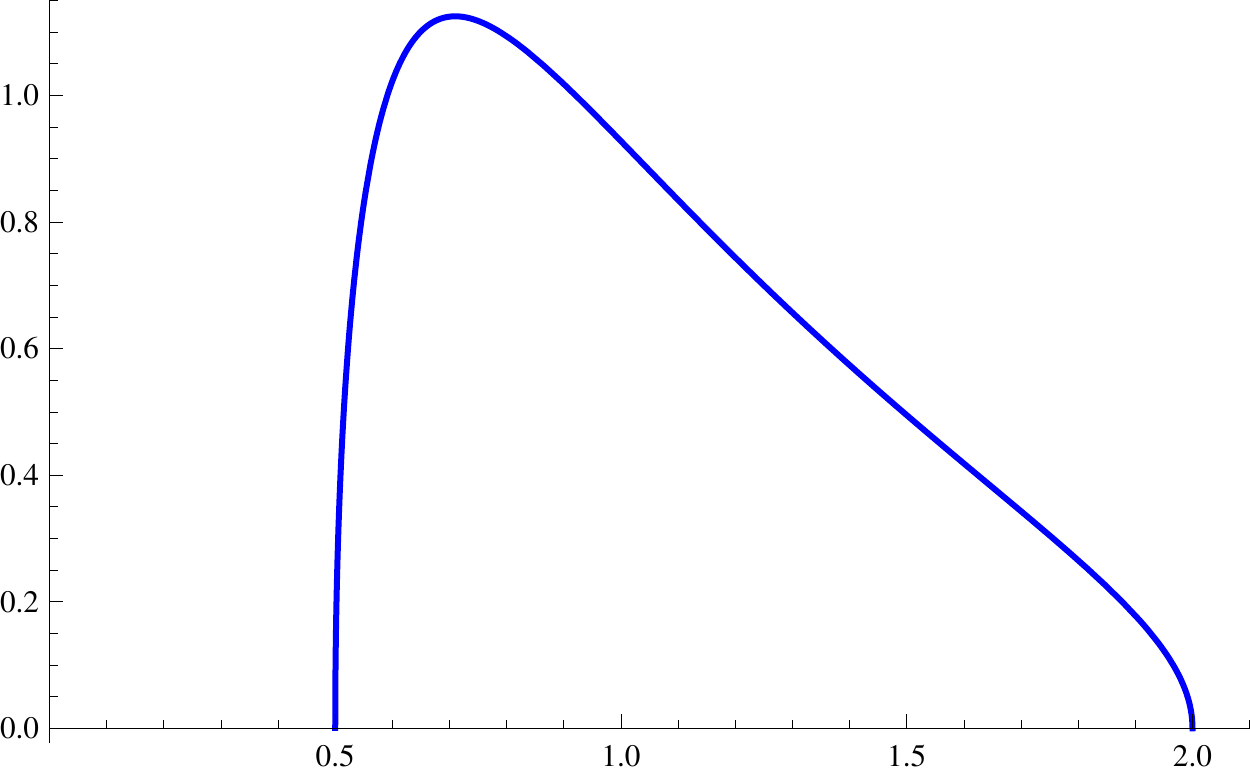} }
\caption{  The  density of eigenvalue (\ref{rho}) for $a=1/2$ in the $x$ plane (left) and in the $z=\re^{x}$ plane (right) .}
 \label{plotrho}
  \end{center}
\end{figure}

Finally, we derive an explicit expression for the density of eigenvalues. From the standard discontinuity equation, we have
\be 
\rho (z)=-{1\over 2  \pi \ri}\left. {v(z) \over 2 \lambda z} \right \vert^{z+\ri 0}_{z-\ri 0} 
={1\over 2 \pi \ri} \left. {v(w) \over 2  \lambda a \,  \sn (w)}\right \vert^{\kk-\ri w }_{\kk+\ri w}, 
\ee
where 
\be 
z= a\,  \sn (\kk+\ri w).
\ee
We find the explicit expression
\be
 \label{rho}
 \ba 
& \rho (z(w))={\ri \over 4 a \cd(\ri w)^2 \left(-1+a^2 \cd(\ri w)^2\right) \kk \pi ^2 \lambda} \times \bigg( 2 \left(1-a^4\right) \kk \nd (\ri w)\sd(\ri w) \kkp \\
& + \left( \cd(\ri w)+a^2 \cd^3(\ri w)  \right) \left(i \pi  w+2 \left(-1+a^4\right) \kk \nd(\ri w) \ssc(\ri w) \kkp+2 \kk Z(\ri w) \kkp\right) \\
& -2 a^4 \cd(\ri w)^2 \kk \sn(\ri w) \kkp-2 a^6 \cd^4 (\ri w)\kk \sn(\ri w) \kkp \bigg).
\ea
\ee
It can be checked that
\be   
 \rho (a)= \rho (z(0))=0,\quad  \rho (1/a)= \rho (z( \kkp))=0 , 
 \ee
 as it should. 
  The explicit form of $\rho$ is shown in \figref{plotrho} for $a=1/2$.

\subsection{Strong coupling behavior}

We will now explore the strong coupling limit of the planar and genus one solution found above. Since all the quantities depend on $\lambda$ through $a$--the endpoint of the cut in the $z$ plane-- 
the first thing to do is to find the relation between $a$ and $\lambda$ for $\lambda$ large. It is convenient to define the new variable
\be
\hat \lambda =\lambda+{1\over 8}. 
\ee
This shift is reminiscent of the shift in $-1/24$ which appears in the exact planar solution of ABJM theory \cite{mp,abjm}, and it might be explained along the same lines, i.e. 
it might correspond to a correction to the D-brane charge \cite{bh,ahho}. If so, it would give a very interesting check of the proposed 
dual geometry $\IC^2 \times \left(\IC^2/\IZ_{N_f}\right)$. 

The exact relation (\ref{tex}) indicates that large $\hat \lambda$ corresponds to $a\rightarrow 0$, and 
the leading order behavior is easily found to be
\be
a \approx \re^{- \pi {\sqrt{2 \hat \lambda}}}, \qquad \hat\lambda  \gg 1. 
\ee
This means that $A$, the endpoint of the cut in the $x$-plane, grows like ${\sqrt{\lambda}}$ for large 't Hooft parameter. This is similar to the behavior in ABJM theory \cite{mp,dmp}. 
It is possible to invert (\ref{tex}) at large $\lambda$, to all orders, and find an expansion 
of the form 
\be
a=2 \,\re^{- \pi {\sqrt{2\hat \lambda}}} \sum_{k=1}^\infty \sum_{\ell=0}^k a_{k,\ell} \re^{- 2k \pi {\sqrt{2 \hat\lambda}}} \hat\lambda^{\ell/2}. 
\ee
For the first few terms, we find
\be
a= 2 \,  \re^{- \pi {\sqrt{2 \hat\lambda}}} \left\{ 1+ 4 \pi {\sqrt{2\hat \lambda}}\re^{- 2\pi {\sqrt{2 \hat\lambda}}} + \left( 80 \pi^2 \hat\lambda -2 -12 \pi  \sqrt{2\hat\lambda }\right) \re^{- 4 \pi {\sqrt{2 \hat\lambda}}} +\cdots\right\}.
\ee
We can use the above results, together with (\ref{free2}) and (\ref{F1EK}), to obtain the expansion of the planar and genus one free energy at strong 't Hooft coupling. We find, for $\lambda \gg 1$, 
\be 
\label{pl-strong}\ba
F_0(\lambda)=& -{ \pi \sqrt{2}\over 3} \hat \lambda^{3/2}+ c + F^{\rm WS}_0 (\lambda), \\
F_1(\lambda)=& \frac{\pi  \sqrt{ 2 \hat \lambda}}{4}-\frac{\log (2 \hat \lambda)}{4}-\frac{1}{3} 2 \log (2)-\frac{5 \log (\pi )}{12}+ F^{\rm WS}_1 (\lambda),
\ea
\ee
where $c$ is a constant of integration  which is determined by matching carefully the weak-coupling expansion (\ref{pert-aF}) to the above asymptotic 
expansion. One finds numerically $c \approx 0.0714$\footnote{After the first version of this paper appeared, Hatsuda and Okuyama conjectured in \cite{hatsuda-o} that $c= \left(\log (2)-\left(\zeta(3)/\pi^2 \right) \right)/8$.}. Moreover,  
\be
\label{ws-ex}
 F^{\rm WS}_g (\lambda)=\sum_{k=1}^\infty \sum_{\ell=0}^k f_{k,\ell}^{(g)} \re^{- 2k \pi {\sqrt{2 \hat\lambda}}} \hat\lambda^{\ell/2} 
\ee
is the contribution of non-perturbative corrections. We will refer to it as the worldsheet instanton contribution. 
These corrections should be due, as in ABJM theory, to worldsheet instantons in the type IIA superstring dual. For the first few terms, we find
\be
\label{ws-f0} \ba
F^{\rm WS}_0(\lambda)&=-{\re^{- 2 \pi {\sqrt{2 \hat\lambda}}} \over 4 \pi^2} \left( 1+ 2 \pi {\sqrt{2\hat \lambda}} \right)-{\re^{- 4 \pi {\sqrt{2 \hat\lambda}}} 
\over 32 \pi^2} \left( 7+ 28 \pi {\sqrt{2 \hat\lambda}} + 64 \pi^2\hat \lambda \right)+\cdots, \\
F^{\rm WS}_1(\lambda)&=\left(\frac{4}{3}-\frac{1}{3} \pi  \sqrt{2 \hat\lambda }\right) \re^{- 2 \pi {\sqrt{2 \hat\lambda}}}+\left(\frac{29}{3}  \pi  \sqrt{2 \hat\lambda }-\frac{11}{6}\right) \re^{- 4 \pi {\sqrt{2 \hat\lambda}}}+\cdots.
\ea \ee
  \begin{figure} \begin{center}
 {\includegraphics[scale=0.5]{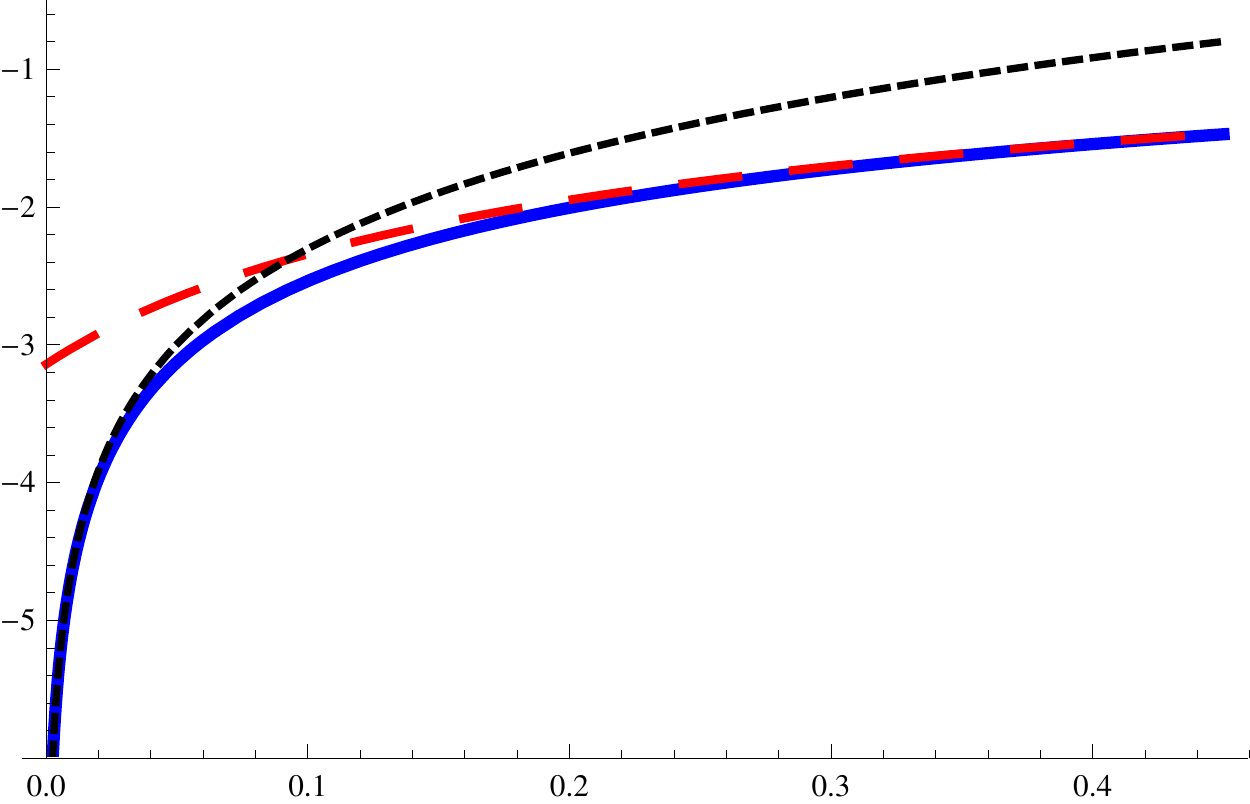}}
\caption{ Comparison of the exact result for $ \rd^2 F_0/ \rd \lambda^2$ given in (\ref{free2}), and plotted in a continuous blue line, with the strong and weak coupling behavior. The red dashed line represents the strong coupling behavior (\ref{pl-strong}), while the black dashed line represents the Gaussian weak coupling behavior (\ref{pert-aF}). }
 \label{plotstwe}
  \end{center}
\end{figure}
Some comments are in order concerning these expressions. First of all, the leading term in (\ref{pl-strong}) has the same form as in ABJM theory, and $N_f$ plays the r\^ole of $k$. This is 
in agreement with the analysis in the strict large $N$ limit, with $N_f$ fixed, performed in \cite{jkps}, and more recently in \cite{mpufu}. However, the structure of the subleading exponential 
terms is different: in ABJM theory, the powers of $\hat \lambda$ appearing in (\ref{ws-ex}) are negative. It would be interesting to understand this in terms of the expansion 
around the conjectural dual worldsheet instantons 
in the dual type IIA theory background. 

\subsection{Grand potential and non-perturbative effects}\label{Gpandnpe}

The model with partition function (\ref{pf}) can be also studied in the Fermi gas approach \cite{mp}. This is a very useful formulation since one can study both the 't Hooft expansion 
and the M-theory expansion, and they lead to two different types of non-perturbative effects. Therefore, the Fermi gas approach makes it possible to go beyond the 
$1/N$ expansion of the matrix model. 

To obtain the Fermi gas picture for the matrix model with partition function (\ref{pf}),  we use the Cauchy identity (\ref{cauchy}) for 
$\mu_i=\nu_i$. We find that (\ref{pf}) can be written as 
\be
Z(N, N_f)={1 \over N!} \sum_{\sigma  \in S_N} (-1)^{\epsilon(\sigma)}  \int    \prod_{i=1}^N  {\rd x_i \over 2 \pi } {1\over \left(2 \cosh\left(  {x_i  \over 2}  \right) \right)^{N_f} 
2 \cosh\left( {x_i - x_{\sigma(i)} \over 2 } \right)}. 
\ee
The corresponding kernel is given by 
\be
\label{densityNf}
\rho_{N_f}(x_1, x_2)={1\over 2 \pi } {1\over \left( 2 \cosh  {x_1 \over 2}  \right)^{N_f /2} }  {1\over \left( 2 \cosh {x_2  \over 2} \right)^{N_f/2} } {1\over 
2 \cosh\left( {x_1 - x_2\over 2 } \right)}. 
\ee

In the Fermi gas approach, the basic quantity is the grand potential of the theory, 
rather than the partition function. As noted in \cite{kkn}, the 't Hooft expansion of the partition function (\ref{pf}) leads naturally to a ``genus" expansion of the grand potential, which is of the form
\be
J(\mu, N_f)=\sum_{g=0}^\infty N_f^{2-2g} J_g \left({\mu \over N_f} \right).
\ee
This expansion contains exactly the same information than the 't Hooft expansion of the (canonical) partition function, and it is related to it by the usual 
thermodynamic transform. In particular, the genus zero piece $J_0$ is just given by the Legendre transform of the planar 
free energy: we first solve for $\lambda$, the 't Hooft parameter, in terms of $\mu/N_f$, through the 
equation 
\be\label{chemicalpot}
{\mu \over N_f}=-{\rd F_0 \over \rd \lambda}, 
\ee
and 
\be
\label{jf}
J_0 \left({\mu \over N_f} \right)= F_0(\lambda) - \lambda {\rd F_0 \over \rd \lambda}.
\ee
Similarly the genus one grand potential $J_1$  is related to the genus one free energy $F_1$ through a one loop saddle point:
\be J_1\left({\mu \over N_f}\right)=F_1\left({\mu\over N_f}\right)+{1\over 2} \log \left( N_f^2 {\partial^2 _ {\mu }  }J_0\left( {\mu \over N_f} \right)\right)-{1\over 2}\log(2\pi).\ee
Equivalently, since $\lambda$ is in one-to-one correspondence with the endpoint of the cut $a$, and all relevant quantities 
are expressed in terms of $a$, we can express $a$ in terms of $\mu/N_f$. 
\be
a = 2\re^{-{2 \mu \over N_f}} \left(1 + \sum_{n\ge 1} \re^{- {4 n \mu \over N_f}} \sum_{l=0}^n b_{n,l} \left( {\mu \over N_f}\right)^{l} \right), 
\ee
and then plug this in the r.h.s. of (\ref{jf}). One obtains, 
\be
\label{joex} \ba
N_f^2 J_0\left({\mu \over N_f} \right)=&\frac{2 \mu ^3}{3 \pi ^2 N_f}-\frac{\mu  N_f}{8}+ N_f^2 \sum_{\ell=1}^\infty \sum_{m=1}^\ell a_{\ell,m}^{(0)} \left({\mu \over N_f} \right)^m \re^{-4  \ell \mu/N_f},\\
J_1\left({\mu \over N_f} \right)=& \frac{1}{2}{\mu \over N_f}+ \sum_{\ell=1}^\infty \sum_{m=1}^\ell a_{\ell,m}^{(1)} \left({\mu \over N_f} \right)^m \re^{-4  \ell \mu/N_f},
\ea
\ee
where
\be
J_g^{\rm WS}\left({\mu \over N_f} \right)= \sum_{\ell=1}^\infty \sum_{m=0}^\ell a_{\ell,m}^{(g)} \left({\mu \over N_f} \right)^m \re^{-4  \ell \mu/N_f}. 
\ee
For the very first orders we find, 
\be
\label{jows} \ba
J_0^{\rm WS}\left({\mu \over N_f} \right)&=-{1\over \pi^2} \left({1\over 4} + {\mu \over N_f} \right)\re^{-4  \mu/N_f}
+{1\over \pi^2} \left( -{7 \over 32}  + {1\over 4}{\mu \over N_f}-4 \left( {\mu \over N_f} \right)^2 \right)\re^{-8  \mu/N_f}  + \cdots, \\
J_1^{\rm WS}\left({\mu \over N_f} \right)&=\left( -\frac{1}{6}-\frac{2}{3} {\mu \over N_f}\right)\re^{-4{\mu \over N_f}}+\left(\frac{11}{12}-6 {\mu \over N_f}\right)\re^{-8{\mu \over N_f}} + \cdots.
\ea\ee
The exponentially small corrections in $\mu$ in (\ref{jows}) are due to the 
worldsheet instanton contributions to the planar and genus one free energy (\ref{ws-f0}). 
The structure of these corrections is quite different from what is obtained in ABJM theory. In this theory, the r\^ole of $N_f$ is played by $k$, and 
after factoring out an overall factor $k^2$ appearing in genus $0$, one finds a simpler structure for the worldsheet instantons
\be
J_0^{\rm WS}\left({\mu \over k} \right)={1\over 4 \pi^2} \sum_{\ell=1}^\infty N_\ell \, \re^{-4  \ell \mu/k},
\ee
where the coefficients $N_\ell$ are related to the genus zero Gromov--Witten invariants of the non-compact Calabi--Yau local $\IP^1 \times \IP^1$ \cite{hmo}.

The planar limit gives us information about the behavior of the theory when 
$N_f$ large and $N/N_f$ is fixed. In the M-theory regime of the theory, we should take $N$ large and $N_f$ fixed. In this regime, based on the 
results of \cite{mp}, we should expect new non-perturbative effects which are not due to worldsheet instantons, but rather 
to membrane instantons. In order to study the M-theory regime, one has in principle to obtain information about the 
spectrum of the operator (\ref{densityNf}) for finite $N_f$. This is however a difficult problem. 
One can then try to study the grand potential of the theory in some approximation scheme. 
In \cite{mp,cm}, various techniques were developed to understand the small $k$ regime of the ABJM model. 
Since $k$ is essentially the Planck constant of the Fermi gas, this is a WKB approximation However, in the model with density matrix (\ref{densityNf}), 
the Planck constant is fixed and set to $2 \pi$, so in principle we can not use the WKB method. However, it was shown in 
\cite{mp} that the perturbative part in $\mu$ of the grand potential only receives quantum corrections up to next-to-leading order, 
and this was recently used in \cite{mpufu} to calculate it. They obtain:
\be
\label{jp}
J_{\rm p}(\mu)={2\over 3 \pi^2 N_f} \mu^3+ \left( {1\over 2 N_f} - {N_f \over 8}\right) \mu. 
\ee
From the point of view of the 't Hooft expansion, this expression contains information about the genus zero and the genus one pieces of the grand potential:
\be
N_f ^2 J^{(0)}_{\rm p}(\mu/N_f )= {2\over 3 \pi^2 N_f} \mu^3- {N_f \over 8} \mu, \qquad  J^{(1)}_{\rm p}(\mu/N_f )={1\over 2} {\mu \over N_f}. 
\ee
This is in agreement with the result in (\ref{joex}). Notice that, at large $N$ (equivalently, large $\mu$), the leading part of the 
grand potential (\ref{jp}) is the cubic part coming from the planar limit. This means that the M-theory limit agrees with the strong coupling expansion of the planar limit, 
in accord with the planar dominance conjecture of \cite{afh}. 

We are interested in calculating non-perturbative corrections to (\ref{jp}), i.e. corrections which are exponentially small in $\mu$. 
In the model (\ref{pf}), $N_f$ plays the r\^ole of $k$, and one could try to study the regime $N_f\rightarrow 0$. To understand the physical nature of this limit, notice that, for large energies, 
the Hamiltonian corresponding to (\ref{densityNf}) is of the form 
\be
H \approx \log \left(2\cosh {p \over 2}\right) + N_f \log \left(2\cosh {q \over 2}\right), 
\ee
and the limit $N_f \rightarrow 0$ corresponds naively to a free theory. However, this limit leads to an IR volume divergence, 
since particles are no longer confined by the $\log \cosh$ potential. For example, the one-particle partition function is given by
\be
Z_1= {1\over 4 \pi} \int_{-\infty}^\infty {\rd q \over \left(2\cosh {q \over 2}\right)^{N_f}}= {1\over 4 \pi} {\Gamma^2(N_f/2) \over \Gamma(N_f)}=
{1\over \pi N_f}-{\pi N_f \over 24} +{\zeta(3) \over  4 \pi} N_f^2 +\cdots, 
\ee
which diverges as $\CO(N_f^{-1})$ when $N_f \to 0$. We then have to extract the leading term in $1/N_f$. To do this, we rescale $q=x/N_f$ as in \cite{mpufu}. 
 In this way we have an explicit Planck constant in the model, $\hbar= 2 \pi N_f$, but we also introduce an explicit $N_f$ dependence in the Hamiltonian:
\be
H\approx \log \left(2\cosh {p \over 2}\right)+ N_f \log \left(2\cosh {x \over 2 N_f}\right).
\ee
This prevents us from applying the WKB method to this problem. We can still extract though the {\it leading} 
contribution to $J(\mu)$ as $N_f \rightarrow 0$, because, in this limit, 
quantum corrections are suppressed. The Hamiltonian becomes
\be
H \approx \log \left(2\cosh {p \over 2}\right) +{|x| \over 2}, \quad N_f\rightarrow 0. 
\ee
The function (\ref{zell}) becomes, in the limit $N_f \rightarrow 0$, 
\be
Z_\ell \approx  \int {\rd p \rd x \over 2 \pi \hbar } {\re^{-\ell |x|/2} \over \left(2\cosh {p \over 2}\right)^{\ell}} ={1\over (2 \pi)^2 N_f } {\Gamma^2(\ell/2) \over \Gamma(\ell)} {4\over \ell},
\ee
and by using (\ref{jsum}) we find 
\be
\label{j-sum}
J(z) \approx - {1\over  \pi^2 N_f} \sum_{\ell \ge 1} {\Gamma^2(\ell/2) \over \Gamma(\ell)} {(-z)^\ell \over \ell^2}, \quad N_f\rightarrow 0.
\ee
This infinite sum in the r.h.s. of (\ref{j-sum}) can be expressed in terms of hypergeometric functions, 
\be
\label{lownf}
N_f J(z) \approx {z\over \pi} 
   \, _3F_2\left(\frac{1}{2},\frac{1}{2},\frac{1}{2};\frac{3}{2},\frac{3}{2};\frac{z^
   2}{4}\right)-{z^2\over 4 \pi ^2} 
   \, _4F_3\left(1,1,1,1;\frac{3}{2},2,2;\frac{z^2}{4}\right),  \quad N_f\rightarrow 0,
   \ee
  and the derivative w.r.t. $z$ has a simpler expression, 
   \be
   \label{derlow}
   {\partial J \over \partial z} \approx {2 \over \pi^2 z N_f } {\rm arcsin}\left({z\over 2}\right) \left( \pi -  {\rm arcsin}\left({z\over 2}\right) \right), \quad N_f\rightarrow 0.
   \ee
Although the building blocks of the functions appearing in (\ref{lownf}) and (\ref{derlow}) have branch cuts in the complex $z$ plane along the positive real axis, and starting at $z=2$, the 
branch cut at positive $z$ disappears in the final answer. This is as it should be, and in accord with previous examples in \cite{mp,cm}: for a quantum Fermi gas, there 
is no physical source of non-analyticity in the grand potential at large fugacity. One can then make an expansion at $\mu$ large to obtain, 
   \be
   \label{leadj}
N_f J(\mu) \approx {2 \mu^3 \over 3 \pi^2} + {\mu \over 2} + {\zeta(3) \over \pi^2}+ J_0^{\rm np} (\mu), \quad N_f\rightarrow 0,
\ee
where
\be
\label{mem-Nf}
J_0^{\rm np} (\mu)=\sum_{\ell \ge 1} \left( a_\ell \mu\,  \re^{-2 \ell \mu} + b_\ell \re^{-2 \ell \mu}\right)= {1\over \pi^2} \left( 2\mu +1\right) \re^{-2\mu}+ 
 {1\over 8\pi^2} \left( 12\mu -1\right) \re^{-4\mu} +\cdots. 
 \ee
The perturbative part in $\mu$ of (\ref{leadj}) agrees with the leading part of (\ref{jp}), at leading order in $N_f$. We find, in addition, exponentially small corrections in $\mu$. Since 
\be
\mu \approx \pi {\sqrt{ N_f N \over 2} } \approx \pi N_f {\sqrt{\lambda \over 2}},
\ee
these corrections are non-perturbative from the point of view of the 't Hooft expansion, which is an expansion in $1/N_f$ at $\lambda$ fixed. They are presumably due to membrane instantons 
in the M-theory dual. 

\subsection{Fermi gas spectrum}

All the information about the partition function (\ref{pf}) and the corresponding grand potential is encoded in the spectrum of the density matrix (\ref{densityNf}). 
As in the case of ABJM theory \cite{mp}, this density matrix can be regarded as a positive Hilbert--Schmidt kernel and its spectrum, defined by
\be 
\label{Nfint}
\int\limits_{-\infty}^{\infty}\rho_{N_f} (x_1,x_2)\phi_n (x_2)\dd x_2=\re^{-E_n}\phi_n(x_1), \qquad n \ge 0, 
\ee
is discrete. We have ordered it as, 
\be
E_0 < E_1 < \cdots.
\ee
When $N_f=1$, the spectrum of this operator is the same as the spectrum of (\ref{densitymat}) for $k=1$, and one can apply the results 
obtained in \cite{km}. 
Unfortunately, for general $N_f$ it doesn't seem to be possible to obtain analytic results for the eigenvalues $E_n$, or an exact quantization condition 
as in \cite{km}. The leading, large $n$ behavior of $E_n$ 
can be obtained by using the techniques of \cite{mp}, and it can be read immediately from (\ref{jp}). Indeed, since 
we are dealing with an ideal Fermi gas, the grand potential can be computed 
from the quantum volume of phase space ${\rm vol}(E)$ as
\be
\label{jton}
 J(\mu, N_f)={1\over 2\pi N_f}\int_{E_0}^{\infty}{ {\rm vol} (E)  \rd E \over \re^{E-\mu}+1}+\cdots, 
 \ee
 where $E_0$ is the ground state energy, and the $\cdots$ denote subleading corrections which appear when we pass 
 from the discrete sum over eigenvalues to the integration over the volume of phase space. The pertubative part of $\mu$ computed in 
 (\ref{jp}) comes from the polynomial part of ${\rm vol}(E)$, 
 \be
{ {\rm vol}_{\rm p}(E)\over 2\pi N_f }={2\over \pi^2 N_f}E^2 -{1\over 6 N_f  }-{N_f \over 8},
 \ee
 which follows from the general results of \cite{mp}. Using now the WKB quantization condition 
 \be
 {\rm vol}(E_n)=2\pi N_f \left( n+{1\over 2} \right) , \quad n \ge 0, 
 \ee
we find the leading behavior, 
\be 
\label{Elead} E_n ^{(0)}=\pi \left({N_f\over 2}\right)^{1/2} \left(n+{1\over 2}+{N_f \over 8}+{1\over 6 N_f}\right)^{1/2}.
\ee
The non-perturbative corrections to $J(\mu, N_f)$ correspond to non-perturbative corrections to ${\rm vol}(E)$, as shown in detail in \cite{km}. As in ABJM theory, we expect two types of 
non-perturbative corrections, of the form 
\be
\re^{-2 \ell E}, \qquad \re^{-4 \ell E/N_f}, \qquad \ell \ge 1. 
\ee
The first type is due to membrane-type corrections, i.e. to the exponentially small corrections appearing in (\ref{mem-Nf}), which are invisible in the 't Hooft expansion. The second type is 
due to worldsheet instantons, i.e. to the exponentially small terms appearing 
in for example (\ref{jows}). Although we do not have an exact asymptotic expansion for generic $N_f$, as we have in ABJM theory, we have results at small $N_f$ for membrane 
corrections, coming from (\ref{leadj}), as well as results at large $N_f$ for worldsheet instanton corrections, coming from (\ref{jows}). 

Let us first analyze the behavior at large $N_f$. It is clear that, in this regime, the leading exponentially small correction is due to the first worldsheet instanton correction. By using (\ref{jton}) as well 
as the results of \cite{km}, we find that it leads to an exponentially small correction to the quantum volume of phase space of the form, 
\be
{{\rm vol}(E) \over 2\pi N_f}\approx { {\rm vol}_{\rm p} (E)\over 2\pi N_f}+ {4\over \pi^2} E \, \re^{-{4 E/ N_f}}, \qquad  N_f \gg 1. 
\ee
Using the WKB quantization condition, we find a correction to the spectrum of the form
\be
\label{AnsatzE} 
E_n \approx  E_n ^{(0)}-N_f \re^{-4  E_n^{(0)}/N_f}, \qquad  N_f \gg 1. 
\ee
Let us now look at the behavior at $N_f \rightarrow 0$. In this case, the grand potential is given by (\ref{lownf}). The leading non-perturbative correction in (\ref{mem-Nf}) leads to a 
correction to the energy levels of the form, 
 \be 
 E_n \approx  E_n ^{(0)}-{1\over 2 E_n^{(0)}  }\re^{-2E_n^{(0)}}, \qquad N_f \ll 1. 
 \ee  

We will now check some of these analytic results against explicit, numerical calculations of the spectrum. 
By using the techniques of \cite{hmo}, one can easily show that  this integral equation is equivalent to an eigenvalue equation for an infinite dimensional Hankel matrix $M$ with entries 
\be
M_{nm}={1\over 4 \pi 2^{N_f} } \int_{-\infty}^\infty \rd q {\tanh^{n+m} (q/2) \over \cosh^{N_f+2}(q/2)}
={1\over 2 \pi 2^{N_f} } {\Gamma\left({1\over 2} + {m+n\over 2}\right) 
\Gamma\left({N_f\over 2}+1\right) \over \Gamma\left( {3\over 2} + {m+n \over 2} + {N_f \over 2}\right)}, 
\ee
when $m+n$ is even, $m , n \ge 0$, otherwise it vanishes. The energy eigenvalues are obtained by diagonalizing $M_{nm}$. To implement this 
numerically, one truncates the matrix $M$ to an $L \times L$ matrix. The eigenvalues of the truncated matrix, $E_{n,L}$,  will converge to  $E_n$ as $L \rightarrow \infty$. 
In order to improve our numerical approximation we apply Richardson extrapolation to
\be E_{n,L}=E_n+\sum\limits_{i\geq 1}{E_n^{i} \over L^i}, 
\ee
as in \cite{km}. Using this procedure, we have computed the first energy levels for various values of $N_f$. 
We will compare these numerical results with the predictions coming from (\ref{jows}) and (\ref{jp}).
As a first check we can compare the numerical eigenvalues $E_{n}^{\rm num}$  to (\ref{Elead}). The results are shown in Table \ref{Tablelead} for $N_f=2$, 
where we show only the first digits. As expected, 
(\ref{Elead}) becomes increasingly good as $n$ is large. 
 \begin{table} \begin{center}
\begin{tabular}{|c|c|c|}  \hline
 $n$ & $E_n ^{(0)}$&   $E_{n}^{\rm num}$ \\ \hline
 0& 2.867869... &  2.88181542992629... \\ \hline
 1 & 4.253737... & 4.2545915286...  \\ \hline
  2&5.288088...  & 5.28819530714...\\ \hline
 3& 6.150893... & 6.1509118188...\\ \hline
 4&6.906742... & 6.906746362...\\ \hline
 \end{tabular}    
 \caption{The numerical eigenvalues $E_{n}^{\rm num}$ against the theoretical leading order result $E_n^{(0)}$, for  $N_f=2$.} \label{tb1}
  \label{Tablelead}
 \end{center}
 \end{table}

\begin{figure}\begin{center}
{\includegraphics[scale=0.5]{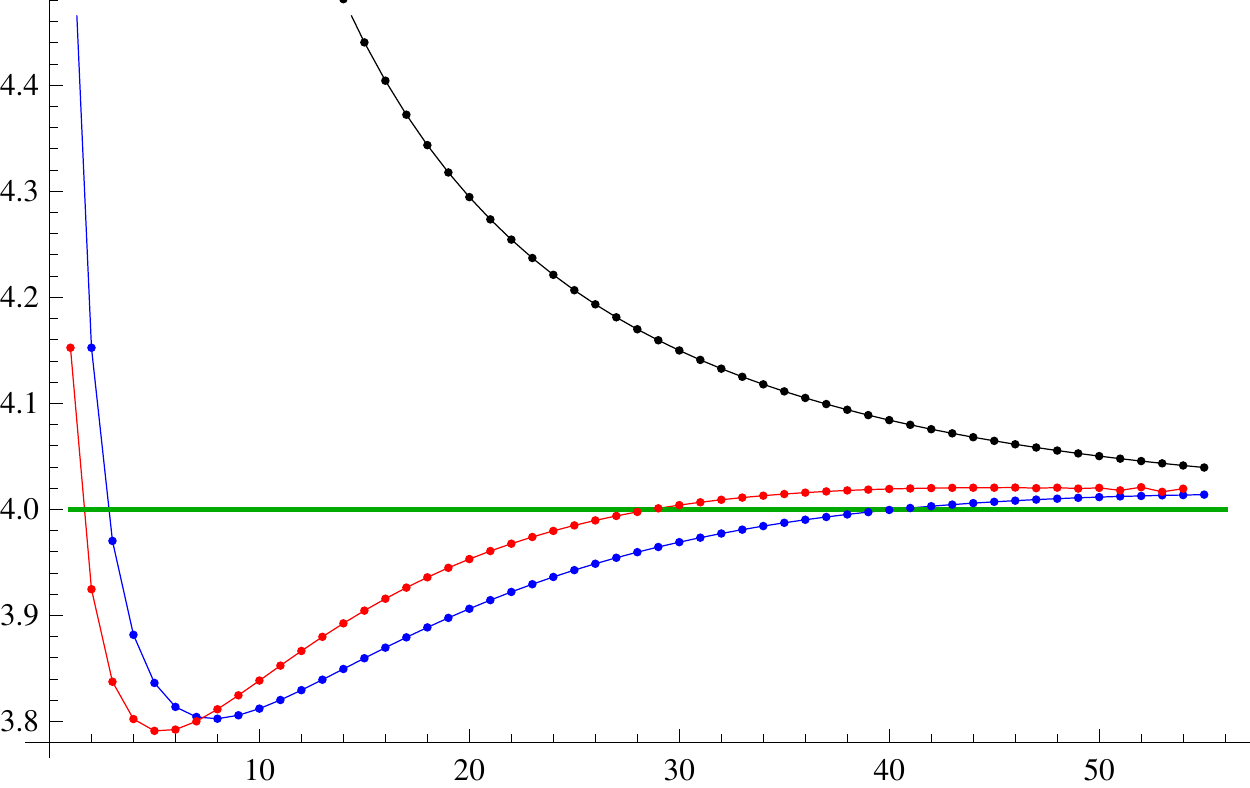}\qquad \qquad \includegraphics[scale=0.5]{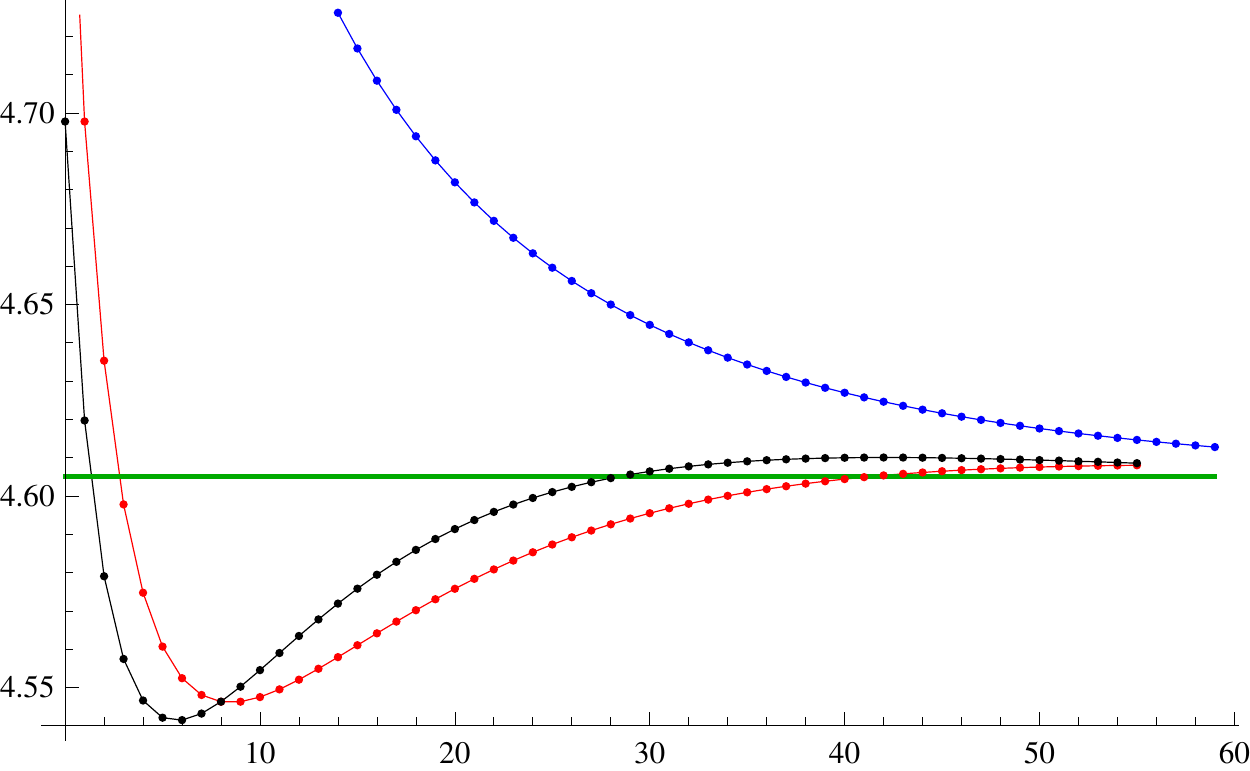}}
\caption{Left: the sequence (\ref{cseq})  for $N_f=100$ with its $3^{\rm th}$ and $4^{\rm th}$ Richardson transform. The straight line is the analytic prediction. Right: 
the sequence (\ref{aseq}) with its $3^{\rm th}$ and $4^{\rm th}$ Richardson extrapolation, again for $N_f=100$. The straight line is the analytic prediction.}
\label{cacoef}
\end{center}
\end{figure}

Next we would like to test exponentially small corrections. The study of the spectrum for small values of $N_f$ is more difficult, since one needs 
very good numerical precision. We will then focus on the study of the large $N_f$, where the dominant non-perturbative correction is (\ref{AnsatzE}). 
This gives a nice numerical verification of the analytical result for the planar limit. To this end we consider the following sequence:
\be 
\label{cseq}  
c_n=-N_f \log \left({E_n-E_n^{(0)}\over  E_{n-1}-E_{n-1}^{(0)}}\right){1\over E_n^{(0)} - E_{n-1}^{(0)}}.
\ee
According to (\ref{AnsatzE}) we should have 
\be c_n \rightarrow  4, \qquad n \rightarrow \infty. 
\ee
This is confirmed by the numerical data, as shown in Figure \ref{cacoef} for $N_f=100$ (we have verified it for other values of $N_f$ as well). 

As a last check of the large $N_f$ behavior, we test the $-N_f$ coefficient of the exponential in (\ref{AnsatzE}). Let us consider the following sequence:
\be \label{aseq}  \log\left(\left| a_n ^{(N_f)} \right| \right)= \log \left(\left| E_n-E_n^{(0)}\right| \right)+{4\over N_f}{E_n^{(0)}} . \ee
According to (\ref{AnsatzE}) we should have 
\be a_n^{(N_f)} \rightarrow  -N_f, \qquad n \rightarrow \infty.
\ee  
This is confirmed by the numerical data, as shown in Figure (\ref{cacoef}) for $N_f=100$. Our best numerical 
approximations, after applying Richardson transformations and taking $n$ sufficiently large, give 
\be   a^{(100)}_n \approx  -100.3, \quad a_n^{(50)}\approx-49.8, \quad a_n^{(40)}\approx -39.6,  \qquad n \gg 1, 
\ee
which compare well to the theoretical value.

\sectiono{The polymer matrix model}

The polymer matrix model is defined by the partition function 
\be
\label{pol}
Z(N, t)={1\over N!}  \int \prod_{i=1}^N {\rd x_i \over 4 \pi}  \re^{- t  \cosh x_i} \prod_{i<j} \left( \tanh \left( {x_i - x_j \over 2} \right) \right)^2. 
\ee
This matrix integral appears in the calculation of two-point functions of the Ising model in two dimensions 
(see for example chapter 20 of \cite{mussardo} for an excellent survey). In this context, the $\tanh$ interaction between 
the eigenvalues is a form factor of order/disorder operators, and the coupling $t$ is interpreted as $mr$, where $m$ is the mass of the Ising fermion 
(proportional to $|T-T_c|$) and $r$ is the Euclidean distance between the two points. 
Let $\Xi$ be the grand canonical partition function associated to the above matrix integral, defined by
\be
\Xi(z, t)= 1+ \sum_{N=1}^\infty Z(N,t) z^N= \Xi_+(z, t) + \Xi_-(z,t), 
\ee
and let 
\be
\Xi_\pm (z,t)={1\over 2} \left( \Xi(z,t) \pm \Xi(-z, t) \right)
\ee
be its even and odd parts w.r.t. $z$. Then, $\Xi_\pm (2, t)$ compute the two-point correlation functions $\langle \mu(r) \mu (0) \rangle$, 
$\langle \sigma(r) \sigma (0) \rangle$ of the disorder operator $\mu(r)$ and the order operator $\sigma(r)$, respectively. 

The matrix integral (\ref{pol}) also plays a r\^ole in a different context: as shown in \cite{fs,zamo}, the grand potential 
\be
J(z, t)= \log \Xi(z,t),
\ee
evaluated at $z=-1$, makes it possible to 
compute the universal scaling functions of a two-dimensional, self-avoiding, non-contractible polymer on a cylinder. This is why we refer to (\ref{pol}) as the polymer matrix model. 
It was also shown in \cite{zamo,twtwo}, based on previous insights in \cite{mtw,cfiv}, that the dependence on the coupling constant $t$ is encoded in an integrable hierarchy of the KdV type, which 
specializes to the sinh-Gordon and the Painlev\'e III equations. 

In all these applications, the large $N$ limit of the integral (\ref{pol}) does not play a crucial r\^ole, since the grand potential has to be 
found at finite values of the fugacity ($z=-1$, $z=2$). However, 
a generalization of the matrix integral (\ref{pol}) can be used to study the six-vertex model on a random lattice \cite{kostov}. 
The planar limit of the model corresponds, as usual, to a lattice of spherical topology. 

As we will see, the polymer matrix model displays a behavior which is very different from the 
one in the ABJM and the $N_f$ matrix models. However, it shares a common feature with them: 
it can be studied in two different regimes, namely the 't Hooft expansion and what we have called an M-theory expansion. 
The 't Hooft expansion is the regime in which 
\be
\label{pol-thooft}
N \rightarrow \infty, \quad t\rightarrow \infty, \qquad \lambda= {N \over t} \quad  \text{fixed}. 
\ee
Indeed, when regarded as an integral over eigenvalues, the polymer matrix model has a $\tanh$ interaction between them and a potential $V(x)=\cosh(x)$. Therefore, 
the coupling $t$ plays the r\^ole of $1/g_s$, and $\lambda$ is then the natural 't Hooft parameter. On the other hand, by using again the Cauchy identity (\ref{cauchy}), 
we can interpret the partition function (\ref{pol}) as the canonical partition function of an ideal Fermi gas with kernel 
\be
\rho(x_1, x_2)={1\over 2 \pi } {\re^{-{t\over2} \cosh(x_1) - {t\over 2}\cosh(x_2)} \over 
2 \cosh\left( {x_1 - x_2\over 2 } \right)}. 
\ee
The Hamiltonian of this gas is, in the leading semiclassical approximation, 
\be
H(x,p)\approx \log \left( 2  \cosh {p \over 2} \right)+ t \cosh(x). 
\ee
The M-theory expansion of (\ref{pol}) is the regime in which 
\be
\label{mpol}
N \rightarrow \infty, \qquad t \, \, \, \text{fixed}, 
\ee
and it corresponds to the thermodynamic limit of the quantum Fermi gas, for a fixed value of the coupling in the 
potential. It is also possible to interpret the matrix integral as a {\it classical} gas with a 
one-body interaction given by $t\cosh(x)$, and a two-body interaction given by $\tanh ((x-x')/2)$. This is the interpretation put forward in \cite{yz,cm,mussardo}. In this interpretation, the 
M-theory expansion corresponds also to the usual thermodynamic limit of the gas.

\subsection{The planar solution}

We will first study the 't Hooft expansion of the polymer matrix model, and we will determine the exact planar limit of the free energy by using similar techniques to those used 
for the $N_f$ matrix model. Indeed, the polymer matrix model is closely related to the $N_f$ matrix model studied in 
detail in the previous section, since the interaction between eigenvalues is the same, 
and only the potential differs. Therefore, after the change of variables $z=\re^x$, we can also regard it as an $O(2)$ model with potential 
\be
V(z) = {1\over 2}(z+ z^{-1}), 
\ee
We can immediately obtain the equations for the endpoints of the cut, by adapting (\ref{o2-end}) to our situation. We find, 
\be\label{ab-cosh}
\ba
{1\over 2 \cos {\pi (1-\nu) \over 2} }  \oint_{C} {\rd z \over 2 \pi \ri} V'(z)  G^{(1-\nu)} (z)&=0 \Rightarrow {1\over 2 \cos {\pi (1-\nu) \over 2} } \left(G^{(1-\nu)}\right)'(0) =1  \\
 {1\over 2 \cos {\pi \nu \over 2} } \oint_{C} {\rd z \over 2 \pi \ri} z V'(z)  G^{(\nu)} (z)&={1\over 2}(2-m)\lambda \Rightarrow -{1\over 2 \cos {\pi \nu \over 2} } G^{(\nu)} (0)+c_1= \lambda(2-m),  
 \ea
 \ee
where $\lambda$ is the 't Hooft parameter defined in (\ref{pol-thooft}), and $c_1$ is defined by the asymptotics
\be
{1\over 2 \cos {\pi \nu \over 2} } G^{(\nu)} (z)={1\over z} + {c_1 \over z^2} + \cdots
\ee
as $z \rightarrow \infty$. 

The first equation is satisfied if $a=1/b$, as in the previous model. The 't Hooft parameter is obtained by studying the limit $\nu \rightarrow 0$ of the second equation. One finds
\be \label{Tcoshex} 
\lambda (a)=\frac{-\pi +\left(2 E(k)+\left(-1+k^2\right) \kk\right) \kkp}{4 k^{1/2} \pi  \kk}, \ee
where 
\be k=a^2.\ee
The planar free energy can be computed by using (\ref{freetau}), and the integration constant can be fixed against 
the behavior near the Gaussian point $\lambda \rightarrow 0$. One finds, 
\be \label{d2Fcosh}
 {\rd^2 F_0 \over \rd \lambda^2}= -{2 \pi \ri \over \tau}.
 \ee 
 By expanding around $\lambda=0$  we get
\be \ba 
\label{wc-pol} a=&1-2 \sqrt{\lambda}+2 \lambda-\frac{\lambda^{3/2}}{2}-\lambda^2+\mathcal{O}(\lambda^{5/2}), \\
 F_0(\lambda)=&\frac{1}{2} \lambda^2 \left( \log\left({\lambda\over 4}\right) -\frac{3 }{2} \right)-\lambda -\frac{\lambda^3}{4}+\frac{5 \lambda^4}{32}-\frac{11 \lambda^5}{64}+\mathcal{O}(\lambda^6).
\ea \ee
This agrees with an explicit perturbative computation. 

\subsection{Strong coupling limit}
We will now look at the 't Hooft expansion the limit $\lambda \rightarrow \infty $. 
The exact relation (\ref{Tcoshex}) indicates that large $\lambda$ correspond to $a \rightarrow 0$. More precisely, one finds
\be 
a(\lambda)\approx {1\over 2 \pi \lambda} W_0\left(\frac{4 \pi  \lambda}{\re} \right), \qquad \lambda \gg 1,
\ee
where $W_0(x)$ is the principal branch of the Lambert function $W(x)$, defined by 
\be
W(x) \re^{W(x)}=x. 
\ee
By using (\ref{d2Fcosh}) this leads to
\be 
F_0(\lambda)\approx {\pi^2 \lambda^2 \over 8 W_0^2\left( \frac{4 \pi  \lambda}{\re} \right)}  \left\{(1-2 \log(2)+2 \log\left[ {1\over 2 \pi \lambda} W_0\left(\frac{4 \pi  \lambda}{\re} \right) \right] \right\}.\ee
By expanding the Lambert function at infinity one gets
\be 
F_0(\lambda) \approx -\frac{\pi ^2   \lambda^2}{4 \log(\lambda)} \left(1+ \mathcal{O}\left({\log\left(\log(\lambda)\right)\over \log (\lambda)}\right) \right).
\ee
Notice that, in this case, the leading order behavior of the free energy is very different from the one appearing in the ABJM matrix model and in the $N_f$ matrix model. 
In addition, we don't have exponentially small corrections in $\lambda$. Of course, in the polymer matrix model we have a very different type of potential, which grows exponentially 
and not linearly, and this leads to a different structure for the free energy. 
 
\subsection{Grand potential and non-perturbative effects}

Let us now analyze the polymer matrix model (\ref{pol}) from the point of view of the Fermi gas. As we reviewed in section \ref{Gpandnpe}, 
the 't Hooft expansion of the matrix model leads to a genus expansion of the grand potential, which now has the 
structure
\be
J(\mu, t)=\sum_{g=0}^\infty t^{2-2g} J_g \left({\mu \over t} \right).
\ee
The relation (\ref{chemicalpot}) in the strong coupling regime leads to
\be\label{muasa} \mu (a)=\frac{\pi  t}{4 a}+ \mathcal{O}(a^3).\ee
Together with (\ref{jf}) this leads to
\be \label{Jth} J_0(\mu)={1\over 2 \pi^2} \left( {\mu \over t}\right)^2 \left(-3- \log\left(\frac{\pi^2 }{64}\right)-2 \log\left(\frac{t}{\mu }\right)\right)
+ \mathcal{O} \left({t^2\over \mu^2} \right). 
\ee

Let us now compare this with a computation in the M-theory limit. As in the other models, working at finite $t$ is difficult, but since $t$ plays the r\^ole of $N_f$ in this model, we can analyze 
the regime in which $t \rightarrow 0$. As we argued before in the case of the $N_f$ model, in this limit we can use the semiclassical approximation. Since
\be
2 \int_0^\infty \re^{-\ell t \cosh q}\, \rd q = 2 K_0 (\ell t) \approx -2 \log(t),  \qquad t\rightarrow 0, 
\ee
where $K_0(z)$ is a modified Bessel function of the second kind, we can calculate $Z_\ell$ as 
\be
Z_\ell \approx -{1\over 2 \pi^2} \log(t) {\Gamma^2 (\ell/2) \over \Gamma(\ell)}, \qquad  t\rightarrow 0. 
\ee
By using (\ref{jsum}), and summing the resulting infinite series, we conclude that 
\be
\label{lpj}
 J(z) \approx 
  { \log(t) \over \pi^2} \left( \arcsin(z/2) - \pi \right) \arcsin(z/2), \qquad t\to 0. 
 \ee
This is a well-known result \cite{mtw,yz}, although the above derivation seems to be simpler 
than the existing ones. In particular, the quantum 
Fermi gas approach to (\ref{pol}) seems to be more 
powerful in obtaining this result than the classical gas approach of \cite{yz,camu}, where one has to treat the interaction term by a Mayer expansion. 
Notice that 
\be
J(2)\approx -{\log(t) \over 4}, \qquad t \rightarrow 0, 
\ee
and 
\be
\Xi(2,t) \approx \left({1\over t}\right)^{1\over 4},  \qquad t \rightarrow 0,
\ee
which is the expected behavior for the correlator of order/disorder operators in the 2d Ising model (see \cite{mussardo}).

 On the other hand, the expression (\ref{lpj}) behaves at large $\mu$ as 
 \be
 J(\mu) \approx  \log\left({1\over t} \right) \left\{ {\mu^2 \over \pi^2} + {1\over 4} + \sum_{\ell\ge 1} \left(a_\ell \mu  +b_\ell\right) \re^{-2 \ell \mu}  \right\}. 
 \ee
 Again, the exponentially small terms at large $\mu$ are invisible in the 't Hooft expansion, and correspond to non-perturbative effects in the M-theory regime of large $N$, small $t$. 

The term (\ref{lpj}) is just the first term in an expansion of $J(z)$ at small $t$ but all orders in $z$. The next terms in this expansion can be computed 
systematically by using the integrable structure of the KdV type underlying the matrix integral (\ref{pol}). The next terms in the expansion have been 
computed in \cite{zamo}:
\be \label{Jzamo} 
J(\mu)={\sigma(\sigma+2)\over 4}\log\left({8\over t}\right) +B(\sigma)+ \mathcal{O} \left( t^{2\pm 2\sigma}\right), \ee
where
\be  
\sigma=- {2\over \pi} \arcsin\left({z\over 2}\right),\qquad B(\sigma)={1\over 4}\int_{0}^{\sigma}\rd x\left(1+x\right)\left[ \psi\left({1+x\over 2}\right)+\psi\left({-1-x\over 2}\right)-2\right].
 \ee
By doing a large $\mu$ expansion of (\ref{Jzamo}) one finds
\be  J(z)=-\mu ^2 \left(\frac{3}{2 \pi ^2}+\frac{\log(\pi )}{\pi ^2}-\frac{\log\left(\frac{8}{t}\right)}{\pi ^2}+\frac{\log\left(\frac{1}{\mu }\right)}{\pi ^2}\right) + \mathcal{O} \left( t^{2\pm 2\sigma}\right)+ \mathcal{O} \left( \mu\log (\mu)\right).
\ee
The  terms of order $\mu^2$  match the result (\ref{Jth}) obtained in the 't Hooft expansion. This is similar to the phenomenon observed in \cite{mp} in the matrix integrals appearing 
in Chern--Simons--matter theories, namely, that the leading, perturbative terms in $\mu$ are the same in both, the 't Hooft expansion and the M-theory expansion. This is again in agreement 
with the planar dominance conjecture of \cite{afh}. 

\sectiono{Conclusions}

In this paper we have studied matrix models which have, on top of the usual 't Hooft regime, an M-theoretic regime. 
These models arise naturally in the localization of Chern--Simons--matter theories 
with M-theory duals, but also in other contexts, like for example the statistical models considered in \cite{kostov-rl,itep}. An important property of these models is that their 
't Hooft expansion is insufficient, and has to be complemented by considering non-perturbative effects which appear naturally in the M-theory regime. 

Our main example has been the matrix model which computes the partition function on the sphere of an $\CN=4$, 3d $U(N)$ gauge theory with one adjoint and $N_f$ fundamental 
hypermultiplets. This theory has a proposed M-theory dual and shares many properties with ABJM theory. We have solved exactly for its planar and genus one limit and started the study of 
its non-perturbative corrections beyond the 't Hooft expansion. A similar model, the polymer matrix model, arises in the study of statistical systems in two dimensions, and we have 
performed a similar analysis. 

The results presented here are just a first step in a more ambitious program which aims at a full understanding of M-theoretic matrix models. In this program, the two 
matrix models which we have studied will probably play an important r\^ole and might be completely solvable, along the lines of the proposed solution of the ABJM matrix model. 
However, it is clear that there are many technical obstacles to face in order to deepen our understanding of M-theoretic matrix models. 
These obstacles were overcome in the study of the ABJM matrix model 
by a series of happy coincidences (mostly, the connection to topological string theory), but cannot be avoided in the more general 
class of models which we would like to study. 

Indeed, one serious drawback of these models is the difficulty to obtain in a realistic way the full 't Hooft 
expansion. It has been shown in \cite{be} that the technique of topological recursion can be in principle 
applied to $O(m)$ models like the one studied in this paper, but in practice it is not easy to apply it (indeed, even for ABJM theory, the 't Hooft 
expansion was obtained in \cite{dmp} by applying the technique of direct integration first proposed in \cite{hk}, and not the topological 
recursion). It is therefore important to develop 
further techniques and ideas to obtain the 't Hooft expansion. 

To understand the M-theoretic regime, we also need to resum the 't Hooft expansion. It is likely that the road to follow here is the one open by the Fermi gas method. 
In order to follow this approach, we should develop techniques to compute the semiclassical expansion of the spectrum of the Fermi gas Hamiltonian, with exponential precision. 
This means that we have to generalize the WKB method to the 
integral equations appearing in this type of problems. As pointed out in \cite{mp,km}, one can obtain in this way a resummed 't Hooft expansion, together with membrane-like 
effects, but then quantum-mechanical instanton corrections have to be included, and these are difficult to compute. 

Another important open problem is to understand the membrane-like corrections from the point of view of the 't Hooft expansion. These are, morally speaking, large $N$ instantons 
of the matrix model (see for example \cite{mmreview}), but it is not clear how to make contact between this point of view and the Fermi gas calculation of these effects. This will probably need 
a better understanding of exponentially small corrections in matrix models. 

Coming back to the concrete models studied in this paper, there are clearly some more precise questions that can be addressed. First of all, 
one could try to determine further terms in the 't Hooft expansion, in both the $N_f$ matrix model and the polymer matrix model. 
In this respect, it would be interesting to see if the direct integration technique of \cite{hk} works also for the $O(m)$ model. The non-perturbative study of the polymer matrix model is probably 
very much facilitated by the connection to classical integrable hierarchies, although a detailed study remains to be done. For the $N_f$ matrix model, the preliminary results 
presented in this paper can be extended and deepened in many ways. One could use the TBA approach of \cite{zamo}, combined with the 
results in \cite{py,hmo-exact}, in order to compute the exact values of $Z(N, N_f)$ for fixed values of $N_f$ and high values of $N$. This would lead to a reasonable ansatz 
for the first terms in the large $\mu$ expansion of the grand potential $J(\mu, N_f)$, as in \cite{hmo}, and might be the starting point for a full non-perturbative study of the model. It would be 
also interesting to see if subleading corrections to the $N_f\rightarrow 0$ limit of the grand potential can be computed analytically from the TBA ansatz. Finally, it would be very interesting to study the eigenvalue problem for the integral equation (\ref{Nfint}) in terms of a difference equation, as it was done in \cite{km} for ABJM theory. This might lead to information about the spectrum at finite $N_f$. We hope to report on some on these issues in the near future.

 \section*{Acknowledgements}
We would like to thank Stefano Cremonesi for calling our attention to the $N_f$ matrix model, and we 
thank him, as well as the participants and organizers of the workshop STAL2013, for interesting 
discussions. We would also like to thank 
M. Mezai, G. Mussardo and S. Pufu for useful communications. 
M.M. would like to thank the Banff Center for hospitality during the conference 
``Modern developments in M-theory." This work is supported by the Fonds National Suisse, subsidies 200020-141329
and 200020-137523.

 \appendix 

\sectiono{The function $G(z)$}
\label{gfunction}

\subsection{General properties} 

The function $G(z)$ was introduced in \cite{EK,EK2} as a technical tool to solve the $O(m)$ matrix model 
for general $m$. It satisfies the defining equation
\be
G(z+\ri 0)+G(z-\ri 0)+2 \cos\left( \pi \nu \right)G(-z) =0, 
   \label{homo-appendix}
\ee
and it is holomorphic on the whole $z$-plane except for the interval $[a,b]$, where it has a branch cut. 
We will map the $z$-plane to the $u$-plane through the equation 
\be
z=a\, {\rm sn} (u,k)= a{\vartheta_3 \over \vartheta_2} {\vartheta_1\left( {u \over 2K} \right) \over \vartheta_0 \left( {u \over 2K} \right)}, 
   \label{u-def}
\ee
where 
\be
\label{kab}
k={a \over b}, 
\ee
and $K$ is the elliptic integral of the first kind with argument $k$.  Here, 
\be
\vartheta_a(v)=\vartheta_a\left( v,\tau\right ), \qquad a=0,1,2,3
\ee
are the elliptic theta functions with modulus
\be
\label{tauK}
\tau= \frac{\ri K'}{K}. 
\ee
Our conventions for elliptic functions and theta functions are as in \cite{akhiezer}. In the following we  use the notation $G(u)$ and $G(z)$ interchangeably. The relationship (\ref{u-def}) can be inverted as
\be
u= \int_0^{z/a}\frac{\rd x}{\sqrt{(1-x^2)(1-k^2x^2)}}. 
\ee
The function $G(u)$ is obtained from the function, 
\be
G_+(u)= \frac{\re^{\pi \ri\nu \over 2}G(u)+\re^{-{\pi \ri\nu \over 2}}G(-u)}{2\sin(\pi\nu)}, 
\label{Gp}
\ee
as
\be
G(u) = -\ri\left[ \re^{\pi \ri\nu \over 2}G_+(u)-\re^{-{\pi \ri\nu \over 2}}G_+(-u) \right]. 
\ee
An explicit expression for $G_+(u)$ was found in \cite{EK2} in terms of theta functions.
Let us define, 
\be
\label{hu}
\ba
H_+(u)
&= \frac{\vartheta_1\left(\frac{u-\ri K'}{2K}\right)\vartheta_1\left(\frac{u-\varepsilon}{2K}\right)}{\vartheta_1\left(\frac{u-K}{2K}\right)\vartheta_1\left(\frac{u-(K+\ri K')}{2K} \right)}
       \re^{-\pi \ri(1-\nu)\frac u{2K}} \nonumber \\
&= -\ri\frac{\vartheta_0\left(\frac{u}{2K}\right)\vartheta_1\left(\frac{u-\varepsilon}{2K}\right)}{\vartheta_2\left(\frac{u}{2K}\right)\vartheta_3\left(\frac{u}{2K} \right)}
       \re^{-\pi \ri(1-\nu)\frac u{2K}}.
\ea
\ee
 In going from the first to the second line, we have used various properties of the theta functions. 
This solution differs from the one given in \cite{EK2} by an overall sign, and follows the conventions in \cite{suyama}. The argument $\varepsilon$ is given by 
\be
\varepsilon= \ri \, (1-\nu)K', 
\ee
and we will denote
\be
\label{ef}
e= a\,  \sn (\varepsilon, k). 
\ee
The function $G_+(z)$ is proportional to $H_+(z)$, and satisfies the normalization condition
\be
\lim_{z\to\infty}zG_+(z) = \ri. 
\ee
Notice that this is the normalization condition chosen in \cite{EK}, and it is different from the one chosen in \cite{EK2}. 
One finds, 
\be
\label{gpex}
G_+(z)={\vartheta_2^2 \over a \vartheta_0 \vartheta_0 \left({ \varepsilon \over 2K}\right) }H_+(z). 
\ee
This can be written in a useful form for the limit $\varepsilon =0$, as follows. We have
 \be
 \ba
 \sn (u,k)&= {\vartheta_3(0)\over \vartheta_2(0)} { \vartheta_1 (u/(2 K)) \over \vartheta_0 (u/(2 K))},\\
  \cn (u,k)&= {\vartheta_0 (0)\over \vartheta_2(0)} { \vartheta_2 (u/(2 K)) \over \vartheta_0 (u/(2 K))},\\
   \dn (u,k)&= {\vartheta_0(0)\over \vartheta_3(0)} { \vartheta_3 (u/(2 K)) \over \vartheta_0 (u/(2 K))}.
   \ea
   \ee
Therefore, 
\be
\left( {\vartheta_0 \vartheta_1\over \vartheta_2 \vartheta_3} \right) \left( {u\over 2 K} \right)= \left( {\vartheta_0 \over \vartheta_3} \right)^2 {\sn (u) \over \cn (u) \dn (u)},
\ee
and we can write the limit $\varepsilon \to 0$ of $G_+(u)$ as 
\be
-{\ri \over a} \left( {\vartheta_2 \over \vartheta_0} \right)^2 \left( {\vartheta_0 \vartheta_1\over \vartheta_2 \vartheta_3} \right) \left( {u\over 2 K} \right)
=-{\ri z \over {\sqrt{(z^2-a^2)(z^2- b^2)}}},
\ee
where we used that
\be
   \cn(u,k)= {\sqrt{1 - z^2/a^2}}, \qquad  \dn(u,k)= {\sqrt{1 - k^2 z^2/a^2}},
   \ee
as well as 
\be
\left( {\vartheta_2 \over \vartheta_3} \right)^2 = k ={a \over b}. 
\ee
Using this result, we find
\be
\label{gpz}
 G_+(z)= -{\ri z \over {\sqrt{ (z^2-a^2)(z^2-b^2)}}} {\vartheta_0 \over \vartheta_0 \left( {\varepsilon \over 2 K} \right)} 
  {\vartheta_1 \left( {u -\varepsilon \over 2K} \right) \over \vartheta_1 \left( {u \over 2K} \right) } \re^{-\pi \ri (1-\nu) u/(2K)}.
 \ee
Finally, we note that the function $G_+(z)$ satisfies the product formula
\be
G_+(z)G_+(-z)={ z^2-e^2 \over (z^2-a^2)(z^2-b^2)}. 
\ee

\subsection{Limiting behavior}
In this paper we need to study the limits $\nu\to 0,\, 1$ of the function $G(z)$. 
Let us first study the limit $\nu \rightarrow 1$. For the function $G_+(u)$, one finds, at first order, 
   \be
   \label{exp-G}
   G_+(z)= -\ri {z\over {\sqrt{(z^2-a^2)(z^2-b^2)}}} \left\{1 -\ri (1-\nu) \left( {\pi u \over 2K} +K' { \vartheta'_1  \over \vartheta_1 }\left( {u \over 2K} \right) \right) + \CO\left( (1-\nu)^2\right) \right\}.
   \ee
In order to obtain an explicit expression for the 't Hooft parameter $\lambda$ in (\ref{venp}), we need to do the expansion up to (and including) third order in $(1-\nu)^3$, and evaluate the result at $z=-1$. Using that
 \be
 \sn \left(K + {\ri K' \over 2} \right)={1\over {\sqrt{k}}}, 
 \ee
 we find that, when $b=1/a$ (which is the case in our model), the point $z=1$ corresponds to
 \be
 u_1 = K + {\ri K' \over 2}, 
 \ee
The point corresponding to $z=-1$ is therefore $u_{-1}=-u_1$, since ${\rm sn}$ is an odd function. In evaluating the coefficients of the expansion of $G(z)$ at $z=\pm 1$, it is 
convenient to use the Jacobi zeta function, which is defined as
\be
Z(u) = {\rd \over \rd u} \log \vartheta_0\left( {u \over 2K} \right). 
\ee
It satisfies the two identities,
\be
\label{Z-idents}
\ba
Z(u+v)& = Z(u) + Z(v)- k^2 {\rm sn} u  \, {\rm sn} v  \, {\rm sn} (u+v), \\
Z(u+ \ri K')&= Z(u) -{\ri \pi \over 2K} + \cs u \,  \dn u. 
\ea
\ee
From (\ref{Z-idents}) one deduces, 
\be
\ba
\left( {\vartheta_1'\over \vartheta_1} \right) \left( {1\over 2} + {\tau \over 4}\right)&= {\ri (k-1) \over 2} - {\ri \pi \over 4 K (k)}, \\
\left({\vartheta_1''\over \vartheta_1}  \right) \left( {1\over 2} + {\tau \over 4}\right)&= 1- k -{E(k) \over K(k)} + \left( \left({\vartheta_1'\over \vartheta_1}  \right) \left( {1\over 2} + {\tau \over 4}\right)\right)^2, \\
\left({\vartheta_1'''\over \vartheta_1}  \right) \left( {1\over 2} + {\tau \over 4}\right)&= 2 \ri k (k-1) + 
3\left( {\vartheta_1'\over \vartheta_1} \right)  \left( {1\over 2} + {\tau \over 4}\right)\left({\vartheta_1''\over \vartheta_1} \right)  \left( {1\over 2} + {\tau \over 4}\right)\\
&-2\left(\left( {\vartheta_1'\over \vartheta_1}  \right) \left( {1\over 2} + {\tau \over 4}\right)\right)^3.
\ea
\ee

In the limit $\nu \rightarrow 0$, the quantity $e$ defined in (\ref{ef}) diverges. Indeed, one has that
  \be
  \label{el0}
 a \,  \sn (-\ri \nu K' + \ri K')= {a \over k \sn (\ri \nu K') } \approx {a\over \ri k \nu K'} 
  \ee
where in the first step we have used an standard identity for the Jacobi $\sn$ function. To calculate the limit of $G(z)$ as $\nu \to 0$, we use that \cite{EK}
\be \label{nunup}
G^{(1-\nu)}(z)=-\left(  \re^{\ri \nu \pi /2}g_{+}(z)G^{(\nu)}_+(z) +g_{+}(-z)G^{(\nu)}_+(-z)  \re^{-\ri \nu \pi /2}\right),
\ee
where 
\be 
g_{+}(z)={\sqrt{(z^2-a^2)(z^2-b^2)}+{z\over e}\sqrt{(e^2-a^2)(e^2-b^2)}\over z^2-e^2}.
\ee
The indices $\nu$, $1-\nu$ indicate that the function G should be evaluated for these values of the parameter.
It follows that 
\be
\lim_{{\nu} \to 0}G(z)=-\lim_{{\nu} \to 1}\left(  \re^{\ri \nu \pi /2}g_{+}(z)G_+(z) +g_{+}(-z)G_+(-z)  \re^{-\ri \nu \pi /2}\right), 
\ee
and we can then use the expansion (\ref{exp-G}) around $\nu=1$. 
 
\sectiono{The $O(m)$ model as a multi-trace matrix model}

Since the planar solution to the $O(m)$ matrix model is relatively complicated, it is useful to make 
an independent computation of various planar quantities. Of course one can do a perturbative computation, but it is 
better to have a more systematic approach which captures the planar limit directly. Such an approach is obtained if one 
regards the $O(m)$ model as a multi-trace matrix model. 

\subsection{Multi-trace matrix models}
Let us consider a matrix model for a Hermitian $N \times N$ matrix, $M$, 
\be
Z= {1\over {\rm vol}(U(N))} \int \rd M \, \re^{-V(M)/g_s}, 
\ee
where the potential is of the form 
\be
V(M)= {1\over 2  } \tr M^2 +t  \sum_{k=1}^{\infty} {a_k\over k} \tr M^k +g_s  \sum_{k,l\ge 1} c_{k,l} 
\tr M^k \tr M^l
\ee
and it includes double-trace operators. We have denoted
\be
t= g_s N, 
\ee
and our conventions are as in \cite{mmhouches}. The standard method to study this type of potentials in the planar limit is to use an analogue of the Hartree--Fock approximation 
\cite{daswadia}. In terms of the density of eigenvalues $\rho(z)$, the planar free energy becomes
\be
g_s^{-2} F[\rho]=t^2 \left( -{1\over 2 t } \rho_2 - \sum_{k\ge 1} {a_k \over k} \rho_k - \sum_{k,l\ge 1} c_{k,l} \rho_k \rho_l +\int \rd \lambda 
\rd \mu \rho(\lambda) \rho (\mu) \log |\lambda-\mu| \right),
\ee
where 
\be
\rho_k =\int \rd \lambda \, \rho(\lambda) \lambda^k. 
\ee
The saddle point equation for $\rho$ is obtained by varying w.r.t. $\rho$:
\be
{1\over 2t} x^2 + \sum_{k=1}^{\infty} {a_{k} \over k} x^{k}+ 2 \sum_{k,l} c_{k,l}  \rho_{l} x^{k} = 
2 \int \rd y \rho(y) \log |x-y| +\zeta,
\ee
where $\zeta$ is a Lagrange multiplier. This equation can be written as 
\be
{1\over t} V_{\rm eff}(x)=2   \int \rd y \rho(y) \log |x-y| +\zeta,
\ee
which is the standard equation appearing in Hermitian matrix model, but it involves the ``effective" potential
\be
\label{effpot}
V_{\rm eff}(x)={1\over 2  } x^2 + \sum_{k=1}^{\infty} t {a_{k} \over k} x^{k}+ 2 t  \sum_{k,l} c_{k,l}  \rho_{l} x^{k},
\ee
which can be written as 
\be
V_{\rm eff}(x)={1\over 2  } x^2 +t \sum_{k\ge 1} { \tilde a_k  \over k} x^k, 
\ee
where 
\be
\tilde a_{k} =a_{k} + 2 k \sum_{l\ge 1}c_{k,l}\rho_{l}.
\ee
Therefore, we can solve for the density of eigenvalues by using this potential and then impose self-consistency. 

We will restrict ourselves to even potentials. In this case, $a_l=0$ for $l$ odd, and 
\be
c_{k,l}=0 \qquad \text{if $k+l=$ odd.}
\ee
This implies that the endpoints of the cut $A,B$ where the 
eigenvalues condense are symmetric $A=-B$. It follows that
\be
\rho_l =0 \qquad l=\text{odd},
\ee
and we have to pick only 
{\it even} terms in the effective potential, i.e. $\tilde a_l=0$ if $l$ is odd. 

We can now treat the effective, even potential with the standard techniques of orthogonal polynomials \cite{biz}. The basic quantity is $R_0(\xi, t)$, which 
can be obtained from the equation
\be
\xi  ={1\over t} R_0 + \sum_{k\ge 1} \tilde a_{2k} {2k-1\choose k-1} R_0^k. 
\ee
The moments $\rho_{2l}$ can be computed as 
\be
\rho_{2l}={(2l)!  \over l!^2} \int_0^1 \rd \xi R_0^l (\xi). 
\ee
In practice we will calculate $\rho_{2l}$ as a power series in $t$:
\be
\rho_{2l} =t^l \sum_{n=0}^{\infty} r_{l,n} t^n, \qquad r_{l,0}= {(2l)!  \over l!^2 (l+1)}.
\ee
We then obtain the following consistency conditions, 
\be
\sum_{n=0}^{\infty} r_{l,n} t^n={(2l)!  \over l!^2} \int_0^1 \rd \xi \Bigl({R_0 (\xi) \over t}\Bigr)^l,
\ee
where 
\be
\label{tildef}
\tilde a_{2k}=a_{2k} + 4 k \sum_{l\ge 1} \sum_{n\ge 0} c_{2k,2l} r_{l,n} t^{n+l}. 
\ee
Since the $\tilde a_k$ are themselves functions of the coefficients $r_{l,n}$, as in (\ref{tildef}), we obtain a set of equations which determine the 
$r_{l,n}$ as functions of $a_k$. This leads to expressions for many of the planar quantities as power series in $t$. If we denote the endpoints of the 
cut as $(-A,A)$, we find
\be
\label{pert-A}
\ba
{A^2 \over 4}=R_0(1)&=t -t^2 a_2 + t^3 \left( a_2^2 -3  a_4-4 c_{2,2}\right) \\
& + 
t^4\left(- a_2^3 + 9  a_2  a_4 -10  a_6 +12 c_{2,2} -8 c_{2,4} -24 c_{4,2} \right)+ \cdots
\ea
\ee
Similarly, the planar free energy can be computed by evaluating the functional $F[\rho]$ on the equilibrium distribution. One easily obtains
\be
F_0(t)=-{t\over 2}  \int \rd x V_{\rm eff} (x) \rho(x) - {t^2\over 2} \zeta +t ^2 \sum_{k,l} c_{k,l} \rho_k \rho_l. 
\ee
Notice that the last term is a correction to the single-trace case. If we use the formalism of orthogonal polynomials, we can rewrite the first 
two terms by using the function $R_0(\xi)$. Our final expression is 
\be
F_0(t)-F^{\rm G}_0(t)=t^2 \int_0^1 \rd \xi (1-\xi) \log \left( {R_0(\xi) \over t \xi} \right)+t ^2 \sum_{k,l} c_{k,l} \rho_k \rho_l,
\ee
where $F^{\rm G}_0(t)$ is the planar free energy of the Gaussian matrix model. 
Like before, this quantity can be computed perturbatively in $t$ in terms of the coefficients of the potential. One finds, 
\be
\label{pert-FF}
F_0(t)-F^{\rm G}_0(t)= -{1\over 2} t^3 a_2 +{1\over 4} t^4 ( a_2^2 - 2  a_4 -4 c_{2,2}) + \CO(t^5).
\ee

\subsection{Examples}

\subsubsection{Chern--Simons matrix model}

The Chern--Simons matrix model describing Chern--Simons theory on $\IS^3$ \cite{mmcs} is a particular case of the 
above multi-trace matrix model \cite{akmv}. In this case, the coefficients are 
given explicitly by the following expressions
\be
a_{2k}= -{2 B_{2k} \over (2k)!}, \quad c_{2k, 2l}= -{ B_{2(k+l)} \over 2(k+l) (2(k+l))!} {2(k+l) \choose 2k},
\ee
where $B_{2k}$ are Bernoulli numbers. Since this model is exactly solvable, we can test the above expressions in detail. For example, (\ref{pert-A}) gives in this case, 
\be
{A^2 \over 4}=t+ \frac{t^2}{6}+\frac{t^3}{90}-\frac{t^4}{2520}-\frac{t^5}{12600}+\CO(t^6),
\ee
which are precisely the first few terms of the perturbative expansion of the exact result
\be
A=2\cosh^{-1}\left( \re^{t/2} \right). 
\ee
The perturbative result (\ref{pert-FF}) for the planar free energy gives
\be
F_0(t)-F^{\rm G}_0(t)={t^3 \over 12} +{t^4 \over 288} -{t^6 \over 86 400}+\CO(t^7),
\ee
which is the expansion of the exact result
\be
F_0(t)-F^{\rm G}_0(t)=-{\rm Li}_3(\re^{-t}). 
\ee
Using the above formalism we can also calculate the correlation functions
\be
W_n(t)=g_s \left\langle  \tr \, \re^{n M}\right \rangle=g_s \sum_{k\ge 0} {n^k \over k!} \tr M^k= g_s \sum_{l\ge 0} {n^{2l} \over (2 l)!} \rho_{2l},
\ee
which correspond to Wilson loops. 

\subsubsection{The $N_f$ and the polymer matrix models}

The matrix models (\ref{pf}), (\ref{pol}) can be written as multi-trace matrix models. In the case of the $N_f$ matrix model, we have
\be
\ba
a_{2k}&= -{4\left(1-2^{2k-1}\right) B_{2k}\over (2k)!} +{1 \over t} (1-\delta_{2,2k}) {4  \left(2^{2k}-1 \right) B_{2k} \over (2k)! } , \\
c_{2k, 2l}&=
-\frac{2 \left(1-2^{2(k+l)-1}\right) B_{2(k+l)} }{(2(k+l)) (2(k+l))!}\binom{2(k+l)}{2k},
\ea
\ee
where  
\be
t= {4 N\over N_f }=4 \lambda. 
\ee
The relative factor of $4$ as compared to (\ref{thooft}) is due to the fact that, in the formalism for multi-trace matrix models developed above, 
the Gaussian potential has the canonical normalization $x^2/2$, while in the expansion of the potential in (\ref{pf}) around $x=0$ we have instead $x^2/8$. 

For the polymer matrix model, we have the same value for $c_{2k,2l}$, but $a_{2k}$ is now, 
\be
a_{2k}=-{4\left(1-2^{2k-1}\right) B_{2k}\over (2k)!} +{1 \over t} (1-\delta_{2,2k}) {1  \over (2k-1)! } . 
\ee
and $t=\lambda$, where $\lambda$ is given in (\ref{pol-thooft}) (the parameter $t$ appearing in (\ref{pol}) should not be confused with the 't Hooft-like parameter $t$ used in this Appendix). 

One should take into account that the coefficients $a_{2k}$ depend now on $1/t$, but it can be easily seen that at each order in $t$ only a finite 
number of terms in the above expansions contribute. After taking these two facts into account, one obtains the results (\ref{pert-aF}), (\ref{wc-pol}), in agreement with 
the exact solution.


\begin{thebibliography}{99}
\bibliographystyle{plain}


\bibitem{kt}
I.~R.~Klebanov and A.~A.~Tseytlin, ``Entropy of near extremal black p-branes,''
  Nucl.\ Phys.\ B {\bf 475}, 164 (1996)
  [hep-th/9604089].
  
   \bibitem{dmp}
 N.~Drukker, M.~Mari\~no, P.~Putrov, ``From weak to strong coupling in ABJM theory,''
  Commun.\ Math.\ Phys.\  {\bf 306}, 511-563 (2011).
  [arXiv:1007.3837 [hep-th]]. 
  
 \bibitem{abjm}
 O.~Aharony, O.~Bergman, D.~L.~Jafferis and J.~Maldacena, ``N=6 superconformal Chern-Simons-matter theories, M2-branes and their gravity duals,''
  JHEP {\bf 0810}, 091 (2008)
  [arXiv:0806.1218 [hep-th]].
  
  \bibitem{bgms}
 S.~Bhattacharyya, A.~Grassi, M.~Mari\~no and A.~Sen, ``A One-Loop Test of Quantum Supergravity,''
  Class.\ Quant.\ Grav.\  {\bf 31}, 015012 (2014)
  [arXiv:1210.6057 [hep-th]].
  
    
\bibitem{mp}
M.~Mari\~no and P.~Putrov, ``ABJM theory as a Fermi gas,''
  J.\ Stat.\ Mech.\  {\bf 1203}, P03001 (2012)
  [arXiv:1110.4066 [hep-th]].

  
   \bibitem{mp1}
M.~Mari\~no and P.~Putrov, ``Exact Results in ABJM Theory from Topological Strings,''
  JHEP {\bf 1006}, 011 (2010)
  [arXiv:0912.3074 [hep-th]].

  
  \bibitem{hmo}
Y.~Hatsuda, S.~Moriyama and K.~Okuyama, 
``Instanton Effects in ABJM Theory from Fermi Gas Approach,''
  JHEP {\bf 1301}, 158 (2013)
  [arXiv:1211.1251 [hep-th]].
  
   \bibitem{hmmo}
 Y.~Hatsuda, M.~Mari\~no, S.~Moriyama and K.~Okuyama, ``Non-perturbative effects and the refined topological string,''
   JHEP {\bf 1409}, 168 (2014)
  [arXiv:1306.1734 [hep-th]].
 
 \bibitem{km}
J.~K\"all\'en and M.~Mari\~no, ``Instanton effects and quantum spectral curves,''
  arXiv:1308.6485 [hep-th].
 
  \bibitem{hmo2}
  Y.~Hatsuda, S.~Moriyama and K.~Okuyama, ``Instanton Bound States in ABJM Theory,''
  JHEP {\bf 1305}, 054 (2013)
  [arXiv:1301.5184 [hep-th]].
  
    \bibitem{mmreview}
 M.~Mari\~no, ``Lectures on non-perturbative effects in large N gauge theories, matrix models and strings,''
  arXiv:1206.6272 [hep-th].
  
    
   \bibitem{hkps}
 C.~P.~Herzog, I.~R.~Klebanov, S.~S.~Pufu and T.~Tesileanu, ``Multi-Matrix Models and Tri-Sasaki Einstein Spaces,''
  Phys.\ Rev.\ D {\bf 83}, 046001 (2011)
  [arXiv:1011.5487 [hep-th]].

\bibitem{itep}
S.~Kharchev, A.~Marshakov, A.~Mironov, A.~Morozov and S.~Pakuliak, ``Conformal matrix models as an alternative to conventional multimatrix models,''
  Nucl.\ Phys.\ B {\bf 404}, 717 (1993)
  [hep-th/9208044].
  
    
 
\bibitem{kostov-rl}  I.~K.~Kostov, ``Solvable statistical models on a random lattice,''
  Nucl.\ Phys.\ Proc.\ Suppl.\  {\bf 45A}, 13 (1996)
  [hep-th/9509124].
  
    \bibitem{abj}
O.~Aharony, O.~Bergman and D.~L.~Jafferis, ``Fractional M2-branes,''
  JHEP {\bf 0811}, 043 (2008)
  [arXiv:0807.4924].
  
    \bibitem{ahs}
H.~Awata, S.~Hirano and M.~Shigemori, ``The Partition Function of ABJ Theory,''
  Prog.\  Theor.\  Exp.\  Phys.\ , 053B04 (2013)
  [arXiv:1212.2966].
  
  \bibitem{honda}
  M.~Honda, ``Direct derivation of "mirror" ABJ partition function,''
  JHEP {\bf 1312}, 046 (2013)
  [arXiv:1310.3126 [hep-th]].
 
 \bibitem{mm}
 S.~Matsumoto and S.~Moriyama, ``ABJ Fractional Brane from ABJM Wilson Loop,''
   JHEP {\bf 1403}, 079 (2014)
  [arXiv:1310.8051 [hep-th]].
  
  \bibitem{suyamaone} 
T.~Suyama, ``On Large $N$ Solution of Gaiotto-Tomasiello Theory,''
  JHEP {\bf 1010}, 101 (2010)
  [arXiv:1008.3950 [hep-th]].
  
  \bibitem{cmp}
 R.~C.~Santamaria, M.~Mari\~no, P.~Putrov, 
 ``Unquenched flavor and tropical geometry in strongly coupled Chern-Simons-matter theories,'' JHEP {\bf 1110} (2011) 139 [arXiv:1011.6281 [hep-th]].
  
    \bibitem{suyama}
 T.~Suyama, ``On Large $N$ Solution of $\CN=3$ Chern-Simons-adjoint Theories,''
  Nucl.\ Phys.\ B {\bf 867}, 887 (2013)
  [arXiv:1208.2096 [hep-th]].

\bibitem{suyamatwo}
  T.~Suyama, ``A Systematic Study on Matrix Models for Chern-Simons-matter Theories,''
  Nucl.\  Phys.\ B {\bf  874}, 528 (2013)
  [arXiv:1304.7831 [hep-th]].
  
  \bibitem{bk}
D.~Bashkirov and A.~Kapustin, ``Supersymmetry enhancement by monopole operators,''
  JHEP {\bf 1105}, 015 (2011)
  [arXiv:1007.4861 [hep-th]].


  \bibitem{bcc}
   F.~Benini, C.~Closset and S.~Cremonesi, ``Chiral flavors and M2-branes at toric CY4 singularities,''
  JHEP {\bf 1002}, 036 (2010)
  [arXiv:0911.4127 [hep-th]].
  
  \bibitem{mpufu}
 M.~Mezei and S.~S.~Pufu, ``Three-sphere free energy for classical gauge groups,''
  JHEP {\bf 1402}, 037 (2014)
  [arXiv:1312.0920 [hep-th], arXiv:1312.0920].
  
    
   \bibitem{kostovon}
  I.~K.~Kostov, ``O($n$) Vector Model on a Planar Random Lattice: Spectrum of Anomalous Dimensions,''
  Mod.\ Phys.\ Lett.\ A {\bf 4}, 217 (1989).
  
\bibitem{ks}
  I.~K.~Kostov and M.~Staudacher, ``Multicritical phases of the $O(n)$ model on a random lattice,''
  Nucl.\ Phys.\ B {\bf 384}, 459 (1992)
  [hep-th/9203030].

  
  \bibitem{EK}
 B.~Eynard and C.~Kristjansen, ``Exact solution of the $O(n)$ model on a random lattice,''
  Nucl.\ Phys.\ B {\bf 455}, 577 (1995)
  [hep-th/9506193].
  
  \bibitem{EK2}
 B.~Eynard and C.~Kristjansen, ``More on the exact solution of the $O(n)$ model on a random lattice and an investigation of the case $|n| > 2$,''
  Nucl.\ Phys.\ B {\bf 466}, 463 (1996)
  [hep-th/9512052].
  
 \bibitem{kwy}
A.~Kapustin, B.~Willett and I.~Yaakov, ``Exact Results for Wilson Loops in Superconformal Chern-Simons Theories with Matter,''
  JHEP {\bf 1003}, 089 (2010)
  [arXiv:0909.4559 [hep-th]].
 
 \bibitem{hama}
N.~Hama, K.~Hosomichi and S.~Lee, ``Notes on SUSY Gauge Theories on Three-Sphere,''
  JHEP {\bf 1103} (2011) 127
  [arXiv:1012.3512 [hep-th]].
  
\bibitem{jafferis}
D.~L.~Jafferis, ``The Exact Superconformal R-Symmetry Extremizes Z,''
  JHEP {\bf 1205}, 159 (2012)
  [arXiv:1012.3210 [hep-th]].
  
\bibitem{jt}
 D.~L.~Jafferis and A.~Tomasiello, ``A Simple class of N=3 gauge/gravity duals,''
  JHEP {\bf 0810}, 101 (2008)
  [arXiv:0808.0864 [hep-th]].
  
   \bibitem{ik}
  Y.~Imamura, K.~Kimura, ``On the moduli space of elliptic Maxwell-Chern-Simons theories,''
  Prog.\ Theor.\ Phys.\  {\bf 120}, 509-523 (2008).
  [arXiv:0806.3727 [hep-th]]. 
  


  
\bibitem{kwytwo}
A.~Kapustin, B.~Willett, I.~Yaakov, ``Nonperturbative Tests of Three-Dimensional Dualities,''
  JHEP {\bf 1010}, 013 (2010).
  [arXiv:1003.5694 [hep-th]].
  
      \bibitem{hmo-exact}
   Y.~Hatsuda, S.~Moriyama and K.~Okuyama, ``Exact Results on the ABJM Fermi Gas,''
  JHEP {\bf 1210}, 020 (2012)
  [arXiv:1207.4283 [hep-th]].
  
  
  \bibitem{dmp-np}
N.~Drukker, M.~Mari\~no and P.~Putrov, ``Nonperturbative aspects of ABJM theory,''
  JHEP {\bf 1111}, 141 (2011)
  [arXiv:1103.4844 [hep-th]].
  
\bibitem{fhm}
H.~Fuji, S.~Hirano and S.~Moriyama, ``Summing Up All Genus Free Energy of ABJM Matrix Model,''
  JHEP {\bf 1108}, 001 (2011)
  [arXiv:1106.4631 [hep-th]].
  
       \bibitem{cm}
 F.~Calvo and M.~Mari\~no, ``Membrane instantons from a semiclassical TBA,''
  JHEP {\bf 1305}, 006 (2013)
  [arXiv:1212.5118 [hep-th]].
  
 \bibitem{py}
  P.~Putrov and M.~Yamazaki, ``Exact ABJM Partition Function from TBA,''
  Mod.\ Phys.\ Lett.\ A {\bf 27}, 1250200 (2012)
  [arXiv:1207.5066 [hep-th]].
  
\bibitem{kkn}
V.~A.~Kazakov, I.~K.~Kostov and N.~A.~Nekrasov, ``D particles, matrix integrals and KP hierarchy,''
  Nucl.\ Phys.\ B {\bf 557}, 413 (1999)
  [hep-th/9810035].

\bibitem{ns}
N.~A.~Nekrasov and S.~L.~Shatashvili, ``Quantization of Integrable Systems and Four Dimensional Gauge Theories,''
  arXiv:0908.4052 [hep-th].

\bibitem{my}
 C.~Meneghelli and G.~Yang, ``Mayer-Cluster Expansion of Instanton Partition Functions and Thermodynamic Bethe Ansatz,''
 JHEP {\bf 1405}, 112 (2014)
  [arXiv:1312.4537 [hep-th]].

\bibitem{bourgine} 
 J.~-E.~Bourgine, ``Notes on Mayer Expansions and Matrix Models,''
  Nucl.\ Phys.\ B {\bf 880}, 476 (2014)
  [arXiv:1310.3566 [hep-th]].

\bibitem{afh}
T.~Azeyanagi, M.~Fujita and M.~Hanada, ``From the planar limit to M-theory,''
  Phys.\ Rev.\ Lett.\  {\bf 110}, no. 12, 121601 (2013)
  [arXiv:1210.3601 [hep-th]].
  
    \bibitem{be}
G.~Borot and  B.~Eynard, ``Enumeration of maps with self avoiding loops and the $O(n)$ model on random lattices of all topologies,''
  J.\ Stat.\ Mech.\ {\bf 01}, 01010 (2001)
  [hep-th/0910.5896].
  
    \bibitem{mmcs}
M.~Mari\~no,
  ``Chern-Simons theory, matrix integrals, and perturbative three-manifold invariants,''
  Commun.\ Math.\ Phys.\  {\bf 253}, 25 (2004)
  [arXiv:hep-th/0207096].

  
\bibitem{tierz}
  M.~Tierz, ``Soft matrix models and Chern-Simons partition functions,''
  Mod.\ Phys.\ Lett.\ A {\bf 19}, 1365 (2004)
  [hep-th/0212128].
  
    \bibitem{hy}
 N.~Halmagyi and V.~Yasnov, ``The Spectral curve of the lens space matrix model,''
  JHEP {\bf 0911}, 104 (2009)
  [hep-th/0311117].
  
       \bibitem{bh}
O.~Bergman and S.~Hirano, ``Anomalous radius shift in AdS$_4$/CFT$_3$,''
  JHEP {\bf 0907}, 016 (2009)
  [arXiv:0902.1743].

  
  \bibitem{ahho}
 O.~Aharony, A.~Hashimoto, S.~Hirano and P.~Ouyang, ``D-brane Charges in Gravitational Duals of 2+1 Dimensional Gauge Theories and Duality Cascades,''
  JHEP {\bf 1001}, 072 (2010)
  [arXiv:0906.2390 [hep-th]].
  
\bibitem{hatsuda-o}
  Y.~Hatsuda and K.~Okuyama, ``Probing non-perturbative effects in M-theory,''
   JHEP {\bf 1410}, 158 (2014)
  [arXiv:1407.3786 [hep-th]].
  
    \bibitem{jkps}
  D.~L.~Jafferis, I.~R.~Klebanov, S.~S.~Pufu and B.~R.~Safdi, 
  ``Towards the F-Theorem: N=2 Field Theories on the Three-Sphere,''
  JHEP {\bf 1106}, 102 (2011)
  [arXiv:1103.1181 [hep-th]].
  
    \bibitem{mussardo}
  G. Mussardo, {\it Statistical Field Theory}, Oxford University Press, Oxford, 2010. 
  
  \bibitem{fs}
P.~Fendley and H.~Saleur, ``N=2 supersymmetry, Painlev\'e III and exact scaling functions in 2-D polymers,''
  Nucl.\ Phys.\ B {\bf 388}, 609 (1992)
  [hep-th/9204094].

      \bibitem{zamo}
  A.~B.~Zamolodchikov, ``Painlev\'e III and 2-d polymers,''
  Nucl.\ Phys.\ B {\bf 432}, 427 (1994)
  [hep-th/9409108].
  
    \bibitem{twtwo}
  C.~A.~Tracy, H.~Widom, ``Fredholm determinants and the mKdV/sinh-Gordon hierarchies," Commun. Math. Phys. {\bf 179}, 1-10 (1996). 

  
       \bibitem{mtw}
 B.~M.~McCoy, C.~A.~Tracy and T.~T.~Wu, ``Painlev\'e Functions of the Third Kind,''
  J.\ Math.\ Phys.\  {\bf 18}, 1058 (1977).

    \bibitem{cfiv}
S.~Cecotti, P.~Fendley, K.~A.~Intriligator and C.~Vafa, ``A new supersymmetric index,''
  Nucl.\ Phys.\ B {\bf 386}, 405 (1992)
  [hep-th/9204102].
  
  \bibitem{kostov}
 I.~K.~Kostov, ``Exact solution of the six vertex model on a random lattice,''
  Nucl.\ Phys.\ B {\bf 575}, 513 (2000)
  [hep-th/9911023].
  
 \bibitem{yz}
 V.~P.~Yurov and A.~B.~Zamolodchikov, ``Correlation functions of integrable 2-D models of relativistic field theory. Ising model,''
  Int.\ J.\ Mod.\ Phys.\ A {\bf 6}, 3419 (1991).
  
  
     \bibitem{camu} 
J.~L.~Cardy and G.~Mussardo, ``Form-factors of Descendent Operators in Perturbed Conformal Field Theories,''
  Nucl.\ Phys.\ B {\bf 340}, 387 (1990).
  
  
  \bibitem{hk}
 M.~x.~Huang and A.~Klemm, ``Holomorphic anomaly in gauge theories and matrix models,''
  JHEP {\bf 0709}, 054 (2007)
  [hep-th/0605195].
  
   \bibitem{akhiezer}
N.I. Akhiezer, {\it Elements of the theory of elliptic functions}, Americal Mathematical Society, Providence, 1990. 

\bibitem{mmhouches}
M.~Mari\~no, ``Les Houches lectures on matrix models and topological strings,''
  hep-th/0410165.
  
   \bibitem{daswadia} 
S.~R.~Das, A.~Dhar, A.~M.~Sengupta and S.~R.~Wadia, ``New critical behavior in $d = 0$ large $N$ matrix models,''
  Mod.\ Phys.\ Lett.\  A {\bf 5}, 1041 (1990).
  
  \bibitem{biz}
D.~Bessis, C.~Itzykson and J.~B.~Zuber,
``Quantum Field Theory Techniques In Graphical Enumeration,''
Adv.\ Appl.\ Math.\  {\bf 1}, 109 (1980).
  
  \bibitem{akmv}
 M.~Aganagic, A.~Klemm, M.~Mari\~no and C.~Vafa, ``Matrix model as a mirror of Chern-Simons theory,''
  JHEP {\bf 0402}, 010 (2004)
  [hep-th/0211098].
  


 


\end{thebibliography}
\end{document}